\documentclass[12pt,english,aps,prd,showpacs,floatfix,headsepline,nofootinbib]{report} %
\usepackage[utf8]{inputenc}
\usepackage{color}
\usepackage{graphicx}
\usepackage{bm}
\usepackage{amsmath}
\usepackage{amsfonts}
\usepackage{eufrak}
\usepackage{hyperref}
\usepackage{lineno}
\usepackage[mathscr]{euscript}
\usepackage{notoccite}

\usepackage[nottoc,notlot,notlof]{tocbibind}

\usepackage{geometry}
\usepackage{setspace}

\newcommand{\mathbit}[1]{\boldsymbol{#1}}

\title{Turbulence in Two-Dimensional Relativistic Hydrodynamic Systems with a Lattice Boltzmann Model}
\author{Mark Watson}

\begin{document}
	
\doublespacing


	\begin{titlepage}
		\begin{center}


			\vspace*{2.5cm}
			 TURBULENCE IN TWO-DIMENSIONAL RELATIVISTIC HYDRODYNAMIC SYSTEMS WITH A LATTICE BOLTZMANN MODEL
			

			\normalsize
			
			\vspace{1.5cm}
			
			by \\
			MARK WATSON \\
			B.S., West Texas A \& M University, 1993 \\
			B.S., West Texas A \& M University, 1996 \\
			M.S., University of Colorado Colorado Springs, 2019 \\
			
			
			\vspace{3.0cm}
			
			A dissertation submitted to the Graduate Faculty of the \\
				University of Colorado Colorado Springs \\
				in partial fulfillment of the \\
			    requirements for the degree of \\
			    Doctor of Philosophy \\
			    Department of Physics and Energy Science \\
		        2022
			
			
			
		\end{center}
\thispagestyle{empty}	
	\end{titlepage}

\pagenumbering{roman}
\setcounter{page}{2}	
	
	\begin{center}
	\vspace*{5.0cm}
	This dissertation for the Doctor of Philosophy degree by \\
	Mark Watson \\
	has been approved for the \\
	Department of Physics and Energy Science \\
	by \\
	
	\vspace{1.0cm}
	
	Paul Romatschke, Chair \\
	Kathrin Spendier \\
	Robert Camley \\
	Michael Calvisi \\
	Mark Hoefer \\
	
	\vspace{1.0cm}
	
		\begin{flushright} 
			Date \underline{\hspace{2em}May 9, 2022\hspace{2em}}
		\end{flushright}
	
	\end{center}
	
\newpage	
	
	
\doublespacing
{\parindent0pt Watson, Mark (Ph.D., Applied Science --- Physics) } \\
{\parindent0pt Turbulence in Two-Dimensional Relativistic Hydrodynamic Systems with a Lattice Boltzmann Model } \\
{\parindent0pt Dissertation directed by Professor Paul Romatschke. }
\begin{center}
	\textbf{ABSTRACT}
\end{center}

Using a Lattice Boltzmann hydrodynamic computational modeler to simulate relativistic fluid systems we explore turbulence in two-dimensional relativistic flows.  We first a give a pedagogical description of the phenomenon of turbulence and its characteristics in a two-dimensional system.  The classical Lattice Boltzmann Method and its extension to relativistic fluid systems is then described.  The model is tested against a system incorporating a random stirring force in k-space and then applied to a realistic sample of graphene.

\textbf{Part I:} We investigate the relativistic adaptation of the Lattice Boltzmann Method reproducing a turbulent, two-dimensional, massless hydrodynamic system with a zero-averaged stirring force randomly generated in momentum space.  The numeric formulation is evaluated and the flow characteristics produced are compared to properties of classical turbulence.  The model can reasonably be expected to offer quantitative simulations of charged fluid flows in two-dimensional relativistic fluid systems.

\textbf{Part II:} At low Reynolds numbers, the wind flow in the wake of a single wind turbine is generally not turbulent.  However, turbines in wind farms affect each other's wakes so that a turbulent flow can arise.  An analogue of this effect for the massless charge carrier flow around obstacles in graphene is outlined.  We use a relativistic hydrodynamic simulation to analyze the flow in a sample containing impurities.  Depending on the density of impurities in the sample, we indeed find evidence for a potentially turbulent flow and discuss experimental consequences.


\tableofcontents

\listoffigures

\newpage

\pagenumbering{arabic}

%
%
\chapter{Introduction}


The study of turbulence in fluid systems is vital to the understanding of dynamics of many fundamental systems in the physical world, from atmospheric weather patterns to astronomical systems, and even the flow of electricity.  Fluid systems with relativistic flows can be found associated with compact astrophysical objects such as the jets accompanying neutron stars or black holes.   Relativistic fluid flows arise naturally in the effects of a collision between two black holes in AdS (Anti-de Sitter) space (see \cite{Bantilan_2018}).  The fluid created by heavy ion collisions, quark-gluon plasma, is of interest for its unique hydrodynamic characteristics \cite{rezzolla_zanotti_2013} \cite{radice2013universality} \cite{Romatschke_2007} \cite{Dusling_2008} \cite{Schenke_2011} \cite{Song_2014} \cite{Bernhard2019} \cite{Mohseni_2014} \cite{weller2017one}.  Work in that fluid system has lead to a number of successful predictions at the RHIC and the LHC.  See, for instance \cite{huovinen2001radial}, \cite{luzum2009viscous}, \cite{habich2015particle} and \cite{romatschke2015light}.  The work in \cite{nagle2014exploiting} successfully predicted the particle emission patterns in 3He + Au collisions at the RHIC. 

Also of interest is the flow of massless charge carriers in a two-dimensional solid lattice such as that formed by graphene, which is considered to be a viscous relativistic charged fluid \cite{2015PhRvB..91h5401F} \cite{Gabbana_2018}.  This system of particles can be viewed as a relativistic fluid since the dynamics of the individual fermions are described by the Dirac equation.  When the particles are considered as a continuous volume they form a fermionic fluid governed by relativistic hydrodynamics.  

Graphene is a single atom thick sheet of graphite with interesting electrical properties.  The two-dimensional solid is made of carbon atoms arranged in a honeycomb lattice, showing a high conductivity through a simple band structure \cite{muller2009graphene} \cite{fritz2008quantum}.  In the charge neutral state each unit in the lattice contains exactly one electron so that half of the energy levels are occupied.  In this state the $sp^2$ hybridized orbit of the electrons in the atoms of the hexagonal lattice form resilient $\sigma$ bonds with neighboring atoms that create a nearly linear dispersion relation, forming so-called Dirac cones on the Fermi surface.  The relation's curves only begin to appear at a very high temperature; $\approx 10^5 K$, validated experimentally in \cite{Zhang2005}\cite{KrishnaKumar2017}.  The band gap disappears at points on the surface at the tips of the cones called Dirac points.  Low energy excitations of the particles and particle holes create highly mobile, massless chiral quasi-particles \cite{Novoselov2005} in the lattice that transport the electric charge through the conical band structure following Dirac's equation of motion \cite{2005Natur.438..197N}.  Near the Dirac points these charge carriers form a relativistic fluid where the Fermi velocity $v_F$ $(\sim 10^6 \frac{m}{s})$ plays the role of the speed of light.  The fluid velocities, bounded from above by $v_F$, are found to be $0.3$ percent the speed of light.  Despite these non-relativistic velocities, the electron fluid behaves very much like a relativistic fluid since there are no mass-scales in the problem. 

The effect of the lack of a mass-scale in the kinetic theory can be seen in the form of the Maxwell-J{\"u}ttner distribution (also called the J{\"u}ttner distribution), which is a probability distribution of the velocities of particles in a hypothetical gas made up of relativistic particles \cite{de1980relativistic}\cite{cercignani2002relativistic}.  Its form is:
\begin{equation} \label{maxwell-juttner-df}
	f(p^\alpha)=D e^{\left( - \frac{U^{\alpha} p_{\alpha}}{k_B T}\right)}.
\end{equation}
In this distribution $U^{\alpha}$ is the fluid 4-velocity, $p_{\alpha}$ is the 4-momentum and $D$ is a normalization constant.  The classical analog of this is the Maxwell-Boltzmann distribution function (also called the Maxwell distribution),
\begin{equation} \nonumber
	f(v) = D e^{\left( - \frac{mv^2}{2 k_B T} \right)} .
\end{equation}
It can be seen that the J{\"u}ttner distribution becomes the Maxwell distribution \cite{aliano2006maxwell} at non-relativistic velocities and non-zero mass.

The transport of the quasi-particles demonstrates a high conductivity that stays above a minimum value, even when carrier concentrations approach zero \cite{Novoselov2005}.  The viscosity to entropy ratio is determined to be very small ($\sim 0.2 \frac{\hbar}{k_B}$)\cite{PhysRevLett.103.025301}; smaller than that of superfluid helium ($\sim 0.8 \frac{\hbar}{k_B}$)\cite{PhysRevLett.103.025301}\cite{bar2014ratio}, and in some cases that of quark-gluon plasma ($0.07 \le \frac{\eta}{s} \le 0.43 \frac{\hbar}{k_B}$)\cite{2013NuPhA.904..377L}.  The number density of the quasi-particles is determined using electrical properties such as the Hall resistance, which normally assumes an integer value in a two-dimensional solid, but in graphene it is found in multiples of $1/2$, known as the fractional Quantum Hall Effect \cite{PhysRevLett.59.1776}.  In the presence of a strong magnetic field the wave functions of the quasi-particles behave like simple quantum harmonic oscillators that take on discrete energy levels based on the strength of the magnetic field.  This effect, called the Shubnikov-De Haas effect, is used to determine the effective mass of the charge carrying quasi-particles.  

Recently, there has been interest in the lattice structure of a solid known as a Kagome solid that also exhibits a linear dispersion relation in its energy spectrum that forms Dirac cones \cite{2019arXiv191106810D}.  This lattice structure appears to support a stronger coupling constant, estimated to be $3.2$ times that of graphene due to the orbital hybridization.  The bulk viscosity caused by phonons and impurities is negligible, so the Coulomb forces dominate and drive a larger fluid velocity with a simultaneously smaller shear viscosity.  These properties allow the emergence of a turbulent massless fluid that is accessible via experiment.  The work by Di Sante et al. \cite{2019arXiv191106810D} used a holographic approach to determine the shear viscosity and Reynolds number of the charged fluid in a Kagome solid.  

Due to the non-linearity of the equations of motion, particularly in the turbulent regime, the study of hydrodynamics often requires a numeric approach to find solutions and many different computational fluid dynamics (CFD) methods have been employed.  The Navier-Stokes solvers include the Finite Volume Method, the Finite Element Method, the Finite Difference Method and the Spectral Element Method.  They solve the Euler and Navier-Stokes equations, and typically perform fairly well.  For open systems, the spectral method ranks high in performance, though the Lattice Boltzmann Method ranks comparable to it for a closed system due to its high parallelizability.  Particle based solvers track individual particles in the fluid and the forces against them.  They include the Lattice Gas Method, Dissipation Particle Dynamics, Multi-Particle Collision Dynamics, Direct Simulation Monte Carlo, and Smooth Particle hydrodynamics.  These solvers provide high accuracy, but large numbers of particles and velocities are a challenge.

The Lattice Boltzmann Method or Lattice Boltzmann Model (LBM) was born out of particle based solvers and is based on the Boltzmann equation.  It is one computational approach that has had a great deal of success in hydrodynamic modeling over the last few decades \cite{PhysRev.94.511}\cite{doi:10.1146/annurev.fluid.30.1.329}\cite{succi2001lattice}\cite{machado2014generalized}\cite{shan1998discretization}\cite{higuera1989lattice}.  It is conceptually simple, highly parallelizable, and unconditionally stable in the presence of shocks owing to its reliance on Boltzmann's probability distribution function.  

LBM modelers of relativistic hydrodynamic flow are less abundant. A Relativistic Lattice Boltzmann Method (RLBM) described in a work from Mendoza, Boghosian, Herrmann and Succi \cite{2010PhRvD..82j5008M} uses parameter matching against conservation laws to fix the form of the equilibrium probability distribution function used in the collision operation. Similar schemes for ultra relativistic flows have been proposed by Mohseni, Mendoza, Succi and Herrmann \cite{2013PhRvD..87h3003M}, and Mendoza, Karlin, Succi and Herrmann \cite{2013PhRvD..87f5027M}.  A full description of a dimension-independent procedure to design a relativistic lattice Boltzmann scheme is presented by Gabbana, Simeoni, Succi and Tripiccione \cite{2020PhR...863....1G}.  A variation of the RLBM proposed by Romatschke, Mendoza and Succi (2011) \cite{2011PhRvC..84c4903R} expands a J{\"u}ttner form of the equilibrium probability distribution function (as opposed to the Maxwellian form used in classical LBM).  An extension of this work, applicable when the chemical potential is not constant, is described by Ambrus and Blaga \cite{2018PhRvC..98c5201A}.  The present work employs another extension of the approach described by \cite{2011PhRvC..84c4903R}.  

In this thesis submission we present the application of a relativistic hydrodynamic numeric modeler to the study of the turbulent characteristics of the unusual flow of massless charge carrying quasi-particles within a two-dimensional solid, using graphene as the principal platform.  

In the first section we begin with a brief description of the classical hydrodynamic equations of motion and their primary moments.  In greater detail we next define and describe the nature of turbulence, how its presence is identified and the differences of the characteristics of turbulence within a two-dimensional vs. a three-dimensional flow.  

In the subsequent section we present a detailed description of the hydrodynamic modeler that is applied in this work, the Lattice Boltzmann Method, including a mathematical justification of its application, and its strengths and weaknesses.  The next two sections describe the application of the modeler to the problem of turbulence.  With the introduction of a forcing term in the relativistic lattice Boltzmann algorithm, we explore the energy propagation resulting from turbulence generated by a two-dimensional stirring force of random magnitude originating in momentum space.  With large enough forcing magnitudes against small viscosity, this force produces Reynolds numbers in the simulation sufficient to create flows in the turbulent regime.  The model is validated through numerical comparison against established turbulent hydrodynamic theory.  Next, we model the flow of the massless quasi-particles in a realistic sample of graphene with one or more obstacles in order to determine if a turbulent signal can be detected.  We conclude with a discussion and general observations of the work as a whole.   

This work describes and validates an effective new formulation of the Lattice Boltzmann Method that can be used to study the effects of turbulence in two-dimensional relativistic systems.  This new methodology is then applied to the charge carrying quasi-particles in graphene and demonstrates that turbulence is possible under specific conditions, even at Reynolds numbers ranging in the transitional phase.

%
%

\chapter{Background}

%
%

	\section{Hydrodynamic Equations of Motion}

              
Hydrodynamics is an effective theory describing the dynamics of a volume of a differentially small fluid sometimes referred to as a parcel.  The objects that constitute the parcel are presumed to be so small and numerous that they are considered collectively as a continuous volume.  As it moves, the boundaries of the parcel can deform and it can collide with other parcels, subject to conservation laws.  The dynamics of the objects that make up the parcel are sacrificed in favor of the local fluid properties; density, pressure, temperature and flow velocity, which are defined locally within the fluid and can vary spatially and temporally throughout \cite{mclean2012continuum}.  These properties are also subject to some thermodynamic equation of state that helps to complete the description of the fluid system.  An excellent derivation of fluid dynamics as a purely continuum theory, without regard to a parcel or its constituent particles is given in \cite{romatschke2019relativistic}.

Taking a Newtonian approach, the fluid's acceleration expressed in terms of the fluid velocity $\bm{u}$ (a three-vector which are indicated in bold) is
\begin{equation} \label{parcel_accel} \nonumber
	\frac{d\bm{u}}{dt} = \frac{\partial \mathbit{u}}{\partial t} + \left( \mathbit{u} \cdot \nabla \right)\mathbit{u}.
\end{equation}
The first term is the local acceleration of the fluid and the second is the advection term that tracks the acceleration of the fluid parcel as it moves from site to site.  

If we assume that forces internal to the fluid vary only spatially within it, and are normal to the parcel boundaries, then we can track an internal driving term as a pressure gradient force, $\bm{F_{int}}=-\frac{\nabla p}{\rho}$, with units of force per unit mass.  It is the difference of the pressure on opposite borders of the parcel (e.g. $\delta p = p_{left} - p_{right}$) times the area of the border (e.g. $\delta y \delta z$), per the mass of the parcel ($m = \rho \delta x \delta y \delta z$).  The acceleration of the parcel with this term for internal forces and some other external forcing term $\bm{g}$ (also in units of force per unit mass) comprise the momentum equation.
\begin{equation} \label{Euler_eqn}
	\frac{\partial \mathbit{u}}{\partial t} + \left( \mathbit{u} \cdot \nabla \right)\mathbit{u} =
	- \frac{\nabla p}{\rho} + \mathbit{g}.
\end{equation}
This is the \emph{Euler equation} for non diffusive and incompressible fluids (i.e $\nabla \cdot \mathbit{u} = 0$) \cite{babinsky2003wings}.

The Navier-Stokes equation of fluid motion adds another forcing term, a dissipative forcing term, employing a viscous stress tensor $\mathscr{T}$.  This tensor is the sum of two spatial derivative expressions of the fluid velocity scaled separately with shear viscosity $\mu$ and bulk viscosity $\lambda$; constants determined experimentally.  Using index notation with Einstein's summation convention, the viscous stress tensor is defined as
\begin{equation} \nonumber
	\tau_{ij} = \mu \left( \frac{\partial u_i}{\partial x^j} + \frac{\partial u_j}{\partial x^i} \right) +
	            \lambda \left(  \frac{\partial u_i}{\partial x^j}  \right) \delta_{ij}.
\end{equation}
The elements of this tensor are presumed to be linear and consistent throughout the fluid, dissipating the fluid based on the different components of the fluid velocity.  They produce a force that is tracked as a divergence across the boundaries of the parcel per unit mass, $\frac{1}{\rho}\nabla \cdot \tau$.  Incorporating this term into \ref{Euler_eqn} gives the \emph{Navier-Stokes equation}:
\begin{equation} \label{Navier_Stokes_eqn}
	\frac{\partial \mathbit{u}}{\partial t} + \left( \mathbit{u} \cdot \nabla \right)\mathbit{u} =
	- \frac{\nabla p}{\rho} + \frac{1}{\rho} \nabla \cdot \mathscr{T} + \mathbit{g}.
\end{equation}
This equation is not generally solvable analytically.  Where possible, solutions are often sought using computational methods.  

For relativistic systems, the dynamics are expressed in space-time components using the energy-momentum 4-tensor $T^{\mu \nu}$ \cite{landau1987course}.  The energy-momentum tensor is the flux of the momentum 4-vector $p^\mu$ over the surface of $x^\nu$.  It contains as its elements the density and current of both energy and momentum. The components are defined by
\begin{equation} \label{stress-energy-tensor-eq}
	T^{\mu \nu} = \left( \epsilon + p \right) u^{\mu} u^{\nu} + g^{\mu \nu} p ,
\end{equation}
where $u^\mu$ is the fluid velocity 4-vector, $\epsilon$ is the energy density, and $p$ is the pressure.  Please note the subtle difference in notation between the scalar pressure $p$ (italics), the momentum 3-vector $\bm{p}$ (boldface), and the momentum 4-vector $p^\mu$ (italics with Greek a upper or lower index).  The tensor $g^{\mu \nu}$ is the space-time metric tensor which, for flat space-time in Minkowski space, has diagonal elements $1, -1, -1$ and $-1$ \cite{landau1987course}.  (Note, \ref{stress-energy-tensor-eq} is expressed in natural units where $c = 1$.)   With this the equations of motion for relativistic fluid dynamics is given simply as
\begin{equation}
	\frac{\partial T_\nu^\mu}{\partial x^\nu} = 0.
\end{equation}
$T^{i0}$ are the spatial componentized momentum densities, $T^{0i}$ are the spatial components of energy current, and $T^{ij}$ are the spatial momentum current components (here $i$, and $j$ are the spatial coordinate indexes $1-3$).  In a co-moving frame of reference (where $u^0 = 0$ and $u^{i}=0$) $T^{00}$ is interpreted physically as the energy density.

Hydrodynamics describes the dynamics of a fluid system relaxing to a thermal equilibrium.  Considering a relatively large volume, and a long enough time scale, the dynamics are limited to locally conserved quantities such as energy, momentum and charge that are determined at the local equilibrium state.  The state principal of equilibrium thermodynamics declares that any local thermodynamic state of the fluid, pressure, density, temperature, internal energy and entropy, can be related to any other two with an equation of state.  When defined, this equation can provide an additional relation to help find solutions to the system.  A common example is the equation of state for an ideal gas \cite{moran2010fundamentals}	; 
\begin{equation} \nonumber
	p = \rho RT \frac{1}{M},	
\end{equation}
where $R$ is the gas constant and $M$ is the molar mass in kilograms per mole.


Ludwig Boltzmann describes the dynamics of a fluid statistically with the Boltzmann transport equation.  For classical systems, it describes the changes of a probability distribution function parameterized with position and momentum across time, $f(t, \mathbit{x}, \mathbit{p})$.  (Velocity is sometimes used in the place of momentum.)  A complete time differential of $f$ is shown in the Liouville expansion
\begin{equation} \nonumber
	\frac{df}{dt} = \frac{\partial f}{\partial t} 
	                + \frac{\bm{p}}{m} \cdot \nabla f 
	                + \bm{F} \cdot \nabla^p f ,
\end{equation}
where $\bm{F} = \frac{d\bm{p}}{dt}$ represents a summation of some external driving forces present causing a change in the momentum, and $(\nabla^p \cdot)$ is the divergence across momentum space.  

The \emph{Boltzmann transport equation} accounts for the dynamics of $f$ by setting it equivalent to a collision term $C[f]$ that represents internal forces caused by interactions between the particles:
\begin{equation} \nonumber
	\frac{\partial f}{\partial t} 
	+ \frac{\bm{p}}{m} \cdot \nabla f 
	+ \bm{F} \cdot \nabla^p f 
	= C[f] .
\end{equation}
The form of the collision operator proposed by Boltzmann is a complicated, multi-volume integral that is computationally challenging even with a numerical approach.  But if we assume collisions are rare the term can simplified.  A popular approximation of this type is the Bhatnagar-Gross-Krook (BGK) collision operator \cite{PhysRev.94.511} \cite{shan1998discretization}.  It assumes that all collisions can be broken down to be between just two particles at a time, and that there are so many collisions that the resulting incidence angles are roughly equally distributed.  Then the local effect of the collisions is to force the quantity $f$ toward an equilibrium.  It represents the collision term as the separation of $f$ locally from its statistical equilibrium $f_{eq}$, taken to have a Maxwellian form classically (see Appendix \ref{chapter:maxwellian_feq_derivation}), with respect to a relaxation time $\tau_{R}$.  
\begin{equation} \label{Boltzmann_equation}
	\frac{\partial f}{\partial t} + \mathbit{u} \cdot \nabla f \frac{d}{dt} + \mathbit{F} \cdot \nabla^{\mathbit{p}} f  
	= \frac{1}{\tau_{R}} \left(f_{eq} - f \right)
\end{equation}
This simplified collision operator ansatz sacrifices the complex microscopic details of the particle interactions, but faithfully preserves the evolution of the fluid's macroscopic moments by maintaining conservation of momentum and energy.  

The integration of the probability function $f$ and another function $g$ over momentum space will yield the average value of $g$; $<g> = \int g f d^3 p$ \cite{2016PhRvA..93a3618B}.  Applying this to orders of the fluid velocity yields expressions for the moments of the fluid; the local mass density $\rho$ (order $0$), the macroscopic fluid velocity $ \bm{u} $ (order $1$), and the total (kinetic plus internal) energy $\epsilon $ (order $2$).
\begin{align} \label{eq:moments_cl}
\nonumber	\rho = \int f(t, \bm{x}, \bm{p}) d^3 \mathbit{p}, \\ 
\nonumber	\rho \mathbit{u} = \int \mathbit{u} f(t, \bm{x}, \bm{p}) d^3 \mathbit{p}, \\
            \epsilon = \int \frac{1}{2} |u|^2 f(t, \bm{x}, \bm{p}) d^3 \mathbit{p}
\end{align}

These relations hold when the flow is at the local equilibrium state as well, where the probability distribution function is the \emph{equilibrium} probability distribution function $f^{eq}$.  Hence, if the macroscopic moments at equilibrium are known, $f^{eq}$ can be determined in terms of them.  The form of $f^{eq}$ can be determined along the same lines of reasoning followed by Maxwell (see Appendix \ref{chapter:maxwellian_feq_derivation}), but it is straightforward to find an expression of the equilibrium probability distribution function in terms of velocity from \ref{eq:moments_cl}.
\begin{equation} \label{eq:eq_dist_funct_cl}
	f^{eq}(t, \bm{x}, \bm{v}) = 
	\frac{\rho}{m} 
	\left( \frac{1}{ 2 c_s^2 (T) \pi } \right) ^{\frac{3}{2}} 
	e^{ \frac{ -\left| \bm{v} - \bm{u} \right|^2}{ 2 c_s^2 (T) } }
\end{equation}
This form introduces the square of the local sound speed as a function of temperature $T$, $c_s^2 (T) = \frac{T}{m}$, and is used within the BGK collision operator to complete the Boltzmann transport equation.  

When close to local equilibrium, the Boltzmann equation can be expanded in powers of the Knudsen number, $Kn = \frac{\tau_R c}{L}$, where $L$ is the characteristic length scale of the system and $\tau_R$ is the relaxation time, which, combined with $c$, represents the mean free path of the particles in the fluid system.  The first term in the expansion yields the Euler equation as follows:  Using the Knudsen number as the choice of smallness parameter $\epsilon$, the probability distribution function can be expanded as
\begin{equation} \nonumber
	f=f^{(0)} + \epsilon f^{(1)} + \epsilon^2 f^{(2)} + ...\text{.}
\end{equation}
The zero order term in the expansion is the equilibrium distribution function $f^{eq}$.  Substituting the zero and first orders into \ref{Boltzmann_equation} (in one dimension for simplicity) we have:
\begin{align} \nonumber
	\frac{\partial}{\partial t} \left( f^{(0)} + \epsilon f^{(1)} \right) +
	u \frac{\partial }{\partial x} \left( f^{(0)} + \epsilon f^{(1)} \right) +
	\frac{F_x}{m} \frac{\partial }{\partial u} \left( f^{(0)} + \epsilon f^{(1)} \right) \\
	= \frac{1}{\tau_R} \left( \left( f^{(0)} + f^{(1)} \right) - \epsilon f^{eq} \right).
\end{align}
Taking only the zero order of $\epsilon$ will give the equilibrium relation
\begin{equation} \nonumber
	\frac{\partial f^{eq}}{\partial t} +
	u \frac{\partial f^{eq} }{\partial x}  +
	\frac{F_x}{m} \frac{\partial f^{eq} }{\partial u} 
	= \frac{1}{\tau_R} \left( f^{eq} - \epsilon f^{eq} \right) = 0.
\end{equation}
Multiply the equation by $u$, integrate over $p$ and apply \ref{eq:moments_cl} to find 
\begin{equation} \nonumber
	\frac{\partial u}{\partial t} +
	u \frac{\partial u }{\partial x}  =
	- \frac{F_x}{m}
\end{equation}
which is the one-dimensional Euler equation.  The second term in the expansion will similarly yield the Navier-Stokes equation.

%
%

	\section{Turbulence}
	
The phenomenon of turbulence has been widely studied, and the literature is numerous.  A few works relevant to this report include (but are not limited to), \cite{migdal2019universal}, \cite{taylor1938spectrum} and \cite{rose1978fully}.  A turbulent flow is characterized by chaotic changes in pressure and velocity throughout the fluid as a result of large kinetic energy in parts of the fluid that overwhelm the viscous damping.  It creates vortices within the flow at different scales; expressed by the curl of the fluid velocity; $\bm{\omega} = \nabla \times \bm{u}$.  The vortex structures cohere to each other, clockwise rotating structures to other clockwise structures, and counterclockwise to counterclockwise, eventually detaching at different scales at periodic intervals, referred to as vortex shedding.  When the periodic intervals cause irregularities in the current, it can potentially be turbulent.

It is difficult to predict quantitatively if turbulence will be present in a fluid system.  The Reynolds number is a unitless ratio of the characteristic length $L$ and average (macroscopic) fluid speed of the system $U$ to the viscous damping term $\nu$; $Re = \frac{LU}{\nu}$.  It is commonly used to identify the boundary between a turbulent and non-turbulent flow.  A large ratio indicates an over-matched damping term, and turbulence is thought to be present when it is larger than something on the order of $5000$.  If the ratio is smaller than that threshold, but larger than $10$, the flow is thought to be in the preturbulent regime.  Any flow with a Reynolds number around or below the order of $10$ is not believed to be turbulent.  It is the most commonly accepted metric to predict the presence of turbulence. 

A consequence of vortex coherence and shedding is a creation and migration of modes in the energy spectrum, also referred to as the power spectrum.  A vortex is formed as some scale in k-space, for example a large scale (small $k$), and new vortices are created at smaller and smaller scales (larger and larger $k$) within the rotating flow until it dissipates at the molecular scale.  This cascading vortex creation is detectable in the energy spectrum $E(k)$ through new modes that appear, having drawn energy from the larger scale modes \cite{berera2018chaotic}.  The energy spectrum is the energy density contained at a given wave number magnitude (denoted here simply as $k$), or the contribution to the energy density by a wave number magnitude \cite{frisch_1995}.
\begin{equation} \nonumber
	E\left( k \right) \equiv \frac{ d \mathcal{E} \left( k \right) }{dk}
\end{equation}  
$ \mathcal{E} \left( k \right) $ is the energy density in spectral space defined as a function of $k$: $ \mathcal{E} \left( k \right) \equiv \frac{1}{2} \langle | \tilde{v} (\bm{k}) |^2 \rangle $, where the brackets represent an average over the angular direction.  The contribution of the total energy density at all scales determines the energy of the system.
\begin{equation}  \nonumber
	\mathcal{E} \left( k \right) = \int_0^{\infty} E\left( k' \right) dk'
\end{equation}

%

The migration of energy to different modes within the energy spectrum, called the cascade of the power spectrum \cite{doi:10.1098/rspa.1991.0075}, or the direct cascade, follows a $-5/3$rd power law.  Kolmogorov \cite{doi:10.1098/rspa.1991.0075} reasoned that the energy would cascade to smaller scales at a rate proportional to the $-5/3$rd power of the wave number.  Explicitly, he calculated through dimensional reasoning that the power spectrum would decay as $ \epsilon^\frac{2}{3}\ k^{-\frac{5}{3}}$ where $\epsilon$ is the energy dissipation rate.  The form of this power law was determined by Kraichnan \cite{1967PhFl...10.1417K}, and has since been observed in simulated and experimental data \cite{2001PhRvE..64c6302D}.  Useful literature describing the cascade of the energy spectrum includes the following: \cite{doi:10.1063/1.1864134}, \cite{merilees_warn_1975}, \cite{scott2007nonrobustness}, \cite{tran2003dual}, \cite{chasnov1997decay}, \cite{tran2002constraints}, \cite{batchelor1969computation} and \cite{fjortoft1953changes}.

Specific to a two-dimensional system, a cascade of the energy spectrum from small scales to larger scales, called the inverse cascade of the energy (or power) spectrum, can also be detected and is also proportional to the $-5/3$rd power of the wave number \cite{doi:10.1098/rspa.1991.0075}\cite{1967PhFl...10.1417K}.  Its presence in a two-dimensional system can be contributed to conservation of enstrophy $Z$, which is the rotational analog of energy and is expressed as $Z=\int_S |\nabla \times \bm{u}|^2$.  It is conserved in two-dimensional systems only.  In a 2D homogeneous, isotropic system, the vorticity equation is \cite{tabeling2002two}:
\begin{equation}
	\frac{\partial \omega }{\partial t} + \bm{u} \nabla \omega = \bm{g} + \nu \Delta \omega
\end{equation}
where $\omega$ is the normal component of the rotation vector $\omega =  \left( \nabla \times \bm{u} \right)_{\hat{\bm{z}}}$, and $\nu$ is the kinematic viscosity.

In three dimensions the viscous strain can cause the rotation vector $\bm{\omega}$ to extend, which can cause vorticity to increase.  In a 2D system, with no forcing term $\bm{g}$ and no viscosity, the vorticity is conserved completely.  Where viscous effects are in play, it can only decrease, and in proportion to the viscosity.  Because of this the cascade of the energy spectrum can only propagate from an initial driving wavenumber to larger scales, resulting in an indirect energy cascade.  However, and enstrophy cascade from large scales to small scales (the direct enstrophy cascade) is possible as the dissipation is directly related to viscosity.  Kolmogorov predicted the log scale slope of this cascade would be $-3$.  That is, the rate of enstrophy dissipation would be proportional to $|k|^{-3}$.  A decent amount of literature on turbulence specific to two-dimensional systems is available, for example the descriptions in the following: \cite{bracco2010reynolds}, \cite{cummins_holloway_2010}, \cite{biskamp1998nonlocal},  \cite{bruneau2005experiments}, \cite{kraichnan1980two} and \cite{boffetta2012two}.


As with the power spectrum, turbulence can be identified through its effect on fluctuations in the spectrum of the current density $\bm{j}$.  When the vortices begin a pattern of coherence and shedding, the flow creates fluctuations in the current density with a defining spectral signal in both frequency space and in wave number space.  When the flow reaches a state of turbulence, the modes spread out from initial prominent modes and the spectrum becomes broadband.  If the broadband signal is present in both frequency space and in k-space the flow is considered turbulent as well as chaotic \cite{HoeferPersonalComm}.  However, if the signal is narrow banded in either space, i.e. the modes do not migrate temporally or spatially, then the flow, though still potentially chaotic, is not considered turbulent.  

\begin{figure}[h!] 
	\centering
	\includegraphics[width=0.85\linewidth]{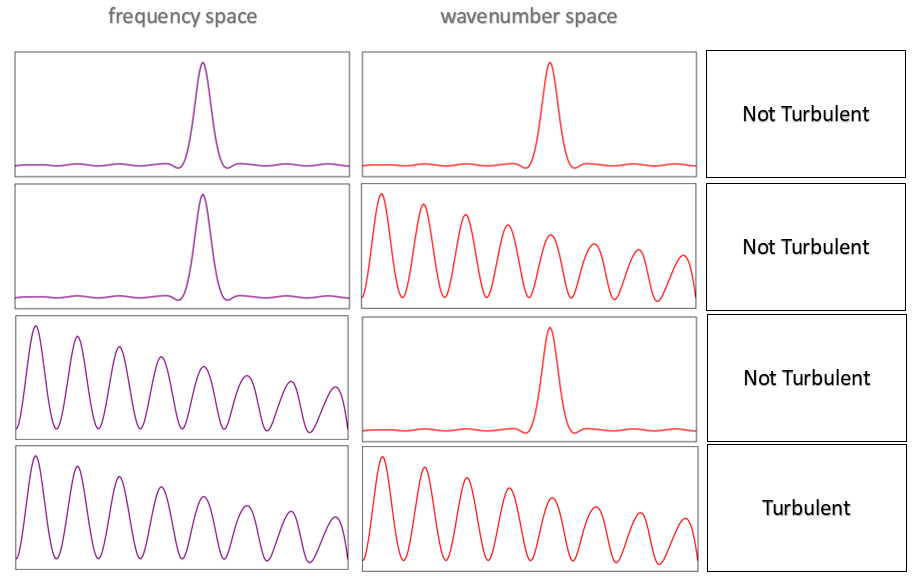}
	\caption[Current density spectra]{A sketch depicting how the spectra of the current density can indicate the presence of a turbulent flow.  If the spectrum in both frequency space and wavenumber space are broadband, the flow can be turbulent.}
	\label{figure:singlemode-broadband}
\end{figure}

%
%
	
	\section{Charged Quasi-Particle Flow}


Of focused interest in this submission is the transport of charge carrying quasi-particles in solid matter.  In particular, we concentrate on the class of two-dimensional solids (solids that are a single atom thick) with an atomic lattice structure that produces a linear dispersion relation in its energy band.  Graphene offers one of the best examples of this class of solid because of its simple atomic structure (a single layer of carbon atoms arranged in a honeycomb structure) and its accessibility to experiment.  It can be wrapped up into zero-dimensional fullerenes, rolled into single-dimensional carbon nanotubes, and be stacked together forming the familiar three-dimensional graphite that can be found in a pencil.  The properties of its band structure were explored by Wallace \cite{WallacePhysRev_71_622} in 1946, but it wasn't until relatively recently that it has been closely examined in experimental settings when it was discovered that it is reasonably simple to produce a useful sample \cite{2005Natur.438..197N}.  

The $sp^2$ hybridized orbit of the electrons in the atoms create resilient bonds between in-plane carbon atoms.  Two electrons in a $p$ orbit and one in an $s$ orbit form $\sigma$ bonds with their neighbors, filling up the shell.  The last $p$ orbit is perpendicular to the plane leaving the shell it occupies half-filled, and the atom charge neutral.  The atoms arrange themselves into a two-dimensional hexagonal pattern with a lattice constant of $a = 1.42 {\AA}$ \cite{RevModPhys.81.109}.  
\begin{figure*}[h!]
	\begin{minipage}{0.48\textwidth} 
		\centering
		\includegraphics[width=0.9\linewidth]{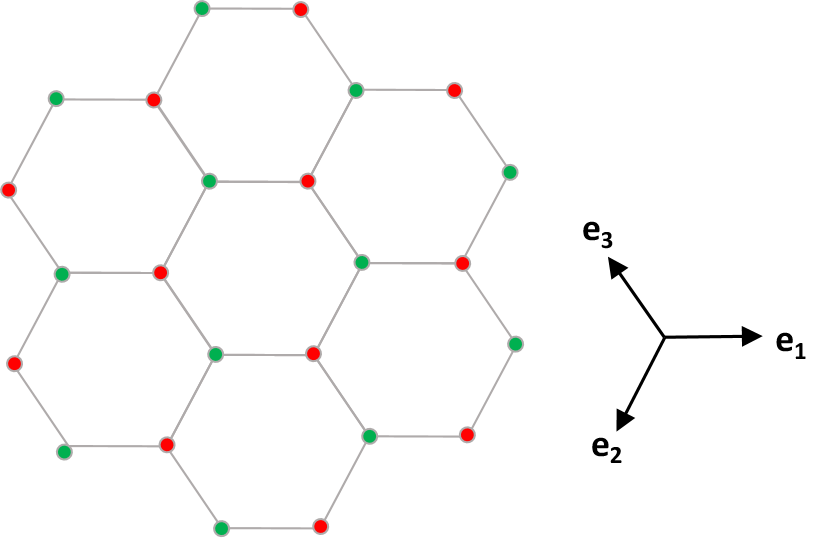}
		\caption[Honeycomb lattice of graphene]{Honeycomb lattice of graphene where the carbon atoms are classified as belonging to either a red or blue sublattice where each carbon atom's nearest neighbor is classified in its companion lattice.}
		\label{red_blue_lattice}
	\end{minipage} \hfill
	\begin{minipage}{0.48\textwidth}
		\centering
		\includegraphics[width=0.7\linewidth]{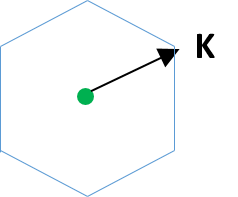}
		\caption[Brillouin zone]{Brillouin Zone in reciprocal space about a carbon atom within a honeycomb lattice of graphene.  The vector $\mathbit{K}$ defines the boundary of the Brillouin zone and, along with the rest of the atoms in the lattice, creates a reciprocal lattice.}
		\label{brillouin_zone}
	\end{minipage}
\end{figure*}	

Wallace determined the band structure of this configuration to be linear, approximated by 
\begin{equation} \label{linear_disp_ref}
	\epsilon(\mathbit{k}) = \pm \hbar v_F |\mathbit{k}|,
\end{equation}
where $v_F \approx 10^6 m/s$ is the Fermi velocity.  This dispersion relation is derived from a `second quantization' approach from the Hamiltonian, given in \cite{doi:10.1142/9789814350525_0009} as
\begin{equation} \label{hubbard_hamiltonian}
	H= - t_1 \sum_{i,j} c_{i\alpha}^{\dagger} c_{j\alpha} 
	 - t_2 \sum_i \left( n_{i\uparrow} + n_{i\downarrow} \right) .
\end{equation}
The constants $t_{1}$ and $t_{2}$ are short range `hopping' amplitudes from a site (indexed by $i$) to the nearest neighbor and next-nearest neighbor, respectively.  The operators $c_{i\alpha}$ and $c_{i\alpha}^{\dagger}$ are the creation and annihilation operators, and the index $\alpha$ indicates the electrons spin (up $\uparrow$, or down $\downarrow$).  The number operators $n_{i\uparrow} \equiv c_{i\uparrow}^{\dagger} c_{i\uparrow}$ and $n_{i\downarrow} \equiv c_{i\downarrow}^{\dagger} c_{i\uparrow}$ are interpreted physically as the electron density. For reference, the commutation relations are as follows,
\begin{align}
	\nonumber \left\{ c_{i\alpha}, c_{j\beta}^{\dagger} \right\}  &= c_{i\alpha} c_{j\beta}^{\dagger} + c_{j\beta}^{\dagger} c_{i\alpha} = \delta_{ij} \delta_{\alpha \beta} \\
	\nonumber \left\{ c_{i\alpha}, c_{j\beta} \right\}  &= c_{i\alpha} c_{j\beta} + c_{j\beta} c_{i\alpha} = 0
\end{align}

With this model the lattice is broken up into two different triangular sub-lattices so that all nearest neighbors of an individual site are part of the companion sub-lattice, as shown in figure \ref{red_blue_lattice} with the two different sub-lattices shown in red and blue.  The basis $e_n$ is chosen in the direction of the nearest neighbors of atoms in the green sub-lattice.
\begin{equation}
	\begin{gathered}
	e_1 = \left(1, 0\right), \ \ \ e_2 = \left( -\frac{1}{2}, \frac{\sqrt{3}}{2}\right), \ \ \ e_2 = \left( -\frac{1}{2}, - \frac{\sqrt{3}}{2}\right) \nonumber \\
	e_i \cdot e_j = - \frac{1}{2} \text{ for } i \ne j, \ \ \ e_1+e_2+e_3 = 0 \nonumber
	\end{gathered} 
\end{equation}
A Fourier transform on \ref{hubbard_hamiltonian} migrates the Hamiltonian to momentum space.  The creation operator transforms as
\begin{align} \nonumber
	c_{A \alpha} \left( \mathbit{k} \right) = 
	\frac{ 1 }{ \sqrt{\mathscr{T}}}  \sum_{i \in A}
	c_{i \alpha} e^{-i \mathbit{k} \cdot \mathbit{e_i}} .
\end{align}
Applying the transformation to the complete form yields
\begin{align} \nonumber
	H =-t_1 \sum_{\mathbit{k}} \left( 
	      e^{i \mathbit{k}\cdot\mathbit{e_1}} + e^{i \mathbit{k}\cdot\mathbit{e_2}} + e^{i \mathbit{k}\cdot\mathbit{e_3}} \right) c_{A\alpha}^{\dagger}(\mathbit{k}) c_{B\alpha}^{\dagger}(\mathbit{k})  + \text{H.c.} \\  
\nonumber  - t_2 \sum_{\mathbit{k}} \left( 
	      c_{A\alpha}^{\dagger}(\mathbit{k}) c_{A\alpha}^{\dagger}(\mathbit{k}) + c_{B\alpha}^{\dagger}(\mathbit{k}) c_{B\alpha}^{\dagger}(\mathbit{k}) 
	      \right) .
\end{align}
The "H.c." term refers to the Hermitian conjugate of the previous term.  This tensor equation is restated using Pauli spin matrices $\tau^x$ and $\tau^y$
\begin{align} \label{hamiltonian_sins_cosines}
	H = \sum_{k} c^{\dagger} (\mathbit{k}) \lbrack[ 
			- t_2 - t_1 \left( cos(\mathbit{k} \cdot e_1) + cos(\mathbit{k} \cdot e_2) + cos(\mathbit{k} \cdot e_3) \right) \tau^x \\ \nonumber
			+ \left(  sin(\mathbit{k} \cdot e_1) + sin(\mathbit{k} \cdot e_2) + sin(\mathbit{k} \cdot e_3) \right) \tau^y
	\rbrack] c (\mathbit{k}).
\end{align}
The energy eigenvalues are found to be 
\begin{equation} \label{energy_eigenvalues} 
	-t_2 \pm | e^{i \mathbit{k}\cdot e_1} + e^{i \mathbit{k}\cdot e_2} + e^{i \mathbit{k}\cdot e_3}  | .
\end{equation}
Diagonalizing \ref{hamiltonian_sins_cosines} yields the dispersion relation  
\begin{equation} \label{full_disp_relation}
	\begin{gathered}
	E_{\pm}(\mathbit{k}) = \pm t_1 \sqrt{3 + f(\mathbit{k})} - t_2 f(\mathbit{k}), \\ 
	f(\mathbit{k}) = 2 cos(\sqrt{3}ak_y) + 4 cos(\frac{\sqrt{3}}{2}ak_x)cos(\frac{\sqrt{3}}{2}ak_y) ,
	\end{gathered}
\end{equation}
where recall $a$ is the distance between the carbon atoms.  

The Brillouin zone is defined in momentum space around an atom where the k-vectors begin to repeat themselves, delineating the lattice cells in reciprocal space.  In graphene it forms a hexagon about the atom, separated from it by a vector $\mathbit{K}$ (fig. \ref{brillouin_zone}) \cite{luo2010elementary}.  If we presume the negligible effects of the next-nearest neighbor on the energy (i.e. $t_2=0$), and we limit the formulation to a single spin (i.e. $\uparrow$), we can expand \ref{full_disp_relation} about a vector $\mathbit{q}$ that is small compared to $\mathbit{K}$ such that $\mathbit{k} = \mathbit{K} + \mathbit{q}$ and $\mathbit{q} \ll \mathbit{K}$ and the dispersion relation is simplified to 
\begin{align} \label{simp_disp_relation}
	E_{\pm}(\mathbit{k}) &\approx \pm t_1 \sqrt{ \left(\frac{3a}{2}q_x\right)^2 + \left(\frac{3a}{2}q_y\right)^2 + O(q_x^3) + O(q_y^3)} \\ \nonumber
	  &= \pm v_F |\mathbit{q}| + O[q^2] ,
\end{align}
where $v_F = \frac{3at_1}{2} \approx 10^6 m/s$.
This dispersion indicates there are two points in the Brillouin zone where the energy is zero.  These are called Dirac points, referred to here as $\mathbit{Q}$ and $-\mathbit{Q}$.  The surface moving away from these points is linear close to the points and forms a conical shape called a Dirac cone \cite{zhen2015spawning} \cite{huynh2011both}.  
\begin{figure}[h!] 
	\centering
	\includegraphics[width=0.5\linewidth]{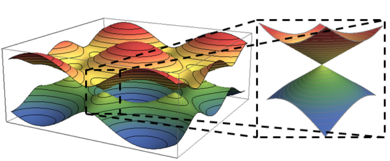}
	\caption[Fermi surface \& Dirac cones]{Fermi surface in the electric band structure of graphene, featuring Dirac cones near the low energy points called Dirac points.}
	\label{figure:Dirac_Cones}
\end{figure}

If we define a continuous Fermi field in the area immediately surrounding the points, we can define an 8 component continuum Fermi field as follows.
\begin{align} \nonumber
	\nonumber C_{A1\alpha}(\mathbit{k}) &= \sqrt{G} c_{A\alpha}\left(\mathbit{Q}_1+\mathbit{k}\right) \nonumber \\ 
	\nonumber C_{A2\alpha}(\mathbit{k}) &= \sqrt{G} c_{A\alpha}\left(-\mathbit{Q}_1+\mathbit{k}\right) \nonumber \\ 
	\nonumber C_{B1\alpha}(\mathbit{k}) &= \sqrt{G} c_{B\alpha}\left(\mathbit{Q}_1+\mathbit{k}\right) \nonumber \\ 
	\nonumber C_{B2\alpha}(\mathbit{k}) &= \sqrt{G} c_{B\alpha}\left(-\mathbit{Q}_1+\mathbit{k}\right) \nonumber \\ 
\end{align}
In this, $G$ is the total area of the honeycomb lattice, and $\mathbit{k}$ is assumed to be small.  The $\alpha$ index refers to the spin, $A$ and $B$ identify the sub-lattice, and the indexes $1$ and $2$ refer to the `valley', or conical area in the Dirac cones.  Again employing Pauli spin matrices for spin $(\sigma^a)$, sub-lattice $(\tau^a)$ and valley $(\rho^a)$ the continuum Hamiltonian is
\begin{equation}  \label{diagonal_continuum_hamiltonian}
	H = \int \frac{d^2k}{4\pi^2} C^\dagger(\mathbit{k}) \left( v_F \tau^y k_x + v_F \tau^x \rho^z k_y \right) C(\mathbit{k})
\end{equation} 
where, again, $v_F$ is the Fermi velocity.  Diagonalizing \ref{diagonal_continuum_hamiltonian} leads to a relativistic spectrum
\begin{equation} \nonumber
	\pm \sqrt{k_x^2 + k_y^2}
\end{equation}
corresponding to the values of \ref{energy_eigenvalues} near the Fermi points.

In the region close to the Dirac points in momentum space where the number density of the electrons approaches zero, one can find long-lived massless quasi particles that carry an electric charge.  The mean free path of the particles is within the fluid regime, and collectively they may be considered a massless, relativistic, viscous charged fluid \cite{2015PhRvB..91h5401F} \cite{Gabbana_2018}.  Note that the velocity of the quasi-particles is not in the relativistic regime, but they are relativistic in the sense that they are massless and their momentum distribution in the fluid is of the J{\"u}ttner form.  Evidence of this viscous charged fluid is noted by \cite{samaddar2021evidence} who identified micrometer sized vortices that cause an inverted electric field which they attribute to a viscous effect.  This field inversion is most prominent near the Dirac points.

%
%

\chapter{The Lattice Boltzmann Method}

    \section{The Standard/Classical LBM}    \label{section:stdLBM}

The standard/classical LBM grew out of the Lattice Gas model, which is a particle-based solver that models particles moving and colliding within a discrete lattice framework.  It replaces the particles in the lattice with the probability distribution function described by the Boltzmann equation (\ref{Boltzmann_equation}), and takes advantage of the near-equilibrium assumption of the BGK collision term, sacrificing collision details for the fluid's macroscopic moments.  The dynamics of the moments are then able to be modeled accurately and efficiently with an appropriately defined computational model.  The Lattice Boltzmann Method or Lattice Boltzmann Model refers to a family of computational methods that model the dynamics of the probability distribution function through the Boltzmann equation projected onto a discrete spatial lattice with a superimposed discrete velocity lattice.  This provides the model with the minimum amount of information required to solve the Boltzmann equation accurately and  efficiently.  There is an abundance of literature on the implementation and application of LBM models; see for instance \cite{succi2001lattice}, \cite{kruger_lattice_2017} \cite{inamuro1995non}, \cite {succi1997lattice}, \cite{TAUZIN2018241} and \cite{wissocq2019extended}.  

In an LBM, the Boltzmann equation is discretized through Chapman-Enskog analysis \cite{chapman1990mathematical} \cite{kumar1967chapman}.  Space is discretized into a spatial lattice where all the information in the probability distribution function for a discrete volume is contained within each node, separated from the other nodes by a lattice spacing in one, two, or three dimensions.  The momentum space is also discrete at every node using a quadrature analysis technique that identifies only the discrete speeds necessary to preserve the moments.  In that way, the LBM lattice can be considered a momentum lattice.  


For each discrete time step the moments of the fluid are evaluated locally at each spatial node.  The moments are obtained through integrals over momentum (or velocity) space such as those defined in \ref{eq:moments_cl}.  The integrals are evaluated with a quadrature integral evaluation technique.  This technique requires dividing momentum space into discrete momenta, which is how the momentum lattice is defined.  Therefore it is the structure that the LBM is built upon.  Considering that, a closer look at quadrature is appropriate.  


For simplicity, consider a one-dimensional function of $x$, $h^N(x)$, that is a non-trivial polynomial of degree $N$, and separable as $h^N(x) = \omega(x)g(x)$.  For this expression $\omega(x)$ is some appropriate weighting function.  An integral of the form $ \int_{-\infty}^{\infty} \omega (x) g(x) dx $ can be approximated by a summation of the function $g(x)$ evaluated at a suitable choice of abscissae $x_i$ and a suitable choice of weights $w_i$ where $i=1...q$ \cite{abramowitzstegun} \cite{abramowitz1948handbook} \cite{deng2010quadrature}.
\begin{equation} \nonumber
	\int_{-\infty}^{\infty} \omega (x) g(x) dx \approx \sum_{i=1}^{q} w_i g(x_i)
\end{equation}
The accuracy of the integration depends on the number of abscissae and how they are chosen.  If we choose the abscissae to be the $n$ roots of the $n$th polynomial $P^{(n)}$ from a class of orthogonal polynomials, and $n \ge \frac{1}{2}(N+1)$, the resulting summation will evaluate the integral \emph{exactly}.    
\begin{equation} \label{eq:gauss_quad_with_ortho_polys}
	\int_{-\infty}^{\infty} \omega (x) g(x) dx = \sum_{i=1}^{n} w_i g(x_i)
\end{equation}
The weights $w_i$ are determined based on the choice of polynomials, which are, in general, chosen based on the limits of integration.  For an integration over $[-1,1]$, Legendre polynomials are useful.  For the range $\left[0,\infty\right)$, Laguerre polynomials are best suited.  And, for the integral limits described in (\ref{eq:gauss_quad_with_ortho_polys}), $(-\infty,\infty)$, Hermite polynomials are the appropriate choice.  



When Hermite polynomials are used the discrete weights are determined by
\begin{equation} \nonumber
	w_i=\frac{n!}{\left(nH^{\left(n-1\right)}\left(x_i\right)\right)^2} ,
\end{equation}
where $H^{(n)}$ is the $n$th order Hermite polynomial.  The Gauss-Hermite quadrature integral relation is generalized to $d$ dimensions \cite{kruger_lattice_2017} as 
\begin{equation}
	\int_{-\infty}^{\infty}{ d^dx \omega\left(\bm{x}\right) g\left(\bm{x}\right) }
	= \sum_{i=1}^{n^d}{ w_i g\left(\bm{x}_i\right) }
\end{equation}

This quadrature technique implies that momentum space can be broken into discrete momenta at each discrete point in space to evaluate the integrals of the fluid moments.  As a simplified example, consider the evaluation of the moments in \ref{eq:moments_cl} with a one-dimensional, classical probability distribution function.
\begin{align}
	\nonumber \rho   =  \int f     dp =  \int w(p) g(p)     dp = \sum_{i} w_i g_i         =& \sum_{i} f_i, \\
	\nonumber \rho u =  \int f v   dp =  \int w(p) g(p) v   dp = \sum_{i} w_i g_i v_i     =& \sum_{i} f_i v_i, \\
	\nonumber \sigma =  \int f v^2 dp =  \int w(p) g(p) v^2 dp = \sum_{i} w_i g_i (v_i)^2 =& \sum_{i} f_i v_i^2
\end{align}

In these summations $f_i$ is probability distribution function for a discrete momentum $i$, determined using quadrature and incorporating the weight.  

The finite summations faithfully reproduce the moment integrals (\ref{eq:moments_cl}) for sufficiently large $n$, implying the discrete abscissae can be used to discretize momentum space.  That is, the roots of the three-dimensional Hermite polynomial of at least the highest order of the moments, the abscissae $x_i$, determine the discrete momenta that are required to connect the nodes of the lattice.  The chosen number of abscissae $n$ need only be one less and half the order of the highest fluid moment; $n=\frac{N-1}{2}$.  Different LBM formulations use different numbers of speeds depending on precision and computational efficiency requirements.  Commonly used values for $n$ with three-dimensional lattices are $19$ and $27$ in configurations named D3Q19 and D3Q27, respectively.  

Using this discretization formula, the continuous probability distribution function is then re-expressed in discrete form: $ f(t,\bm{x},\bm{p}) \rightarrow f_i(t,x_j) $, where $f_i$ is the probability distribution function at lattice node $x_j$ for the discrete speed $i$, incorporating the weighting factor $w_i$. The Lattice Boltzmann Equation is then:
\begin{equation} \label{eq:lbe_cl}
	f_i \left(t + \Delta t, x_j + v_i \Delta t \right) - f_i \left( t, x_j \right) 
	= \Delta t C \left[ f_i \left( t, x_j \right) \right] .
\end{equation}
The collision term on the right-hand side is
\begin{equation} \label{eq:lbe_coll_term}
	C \left[ f_i \left( t, x_j \right) \right] =\frac{1}{\tau_{R}} \left(f^{eq}_i - f_i \right) ,
\end{equation}
where $f^{eq}_i$ is the local equilibrium distribution function for lattice node $i$.  Notice that, with this formulation, the full value of the probability distribution function at each node is never needed.  The value at each discrete momentum is determined independently by the same $f_i$ of neighboring nodes and the equilibrium distribution function $f^{eq}_i$, which is determined by the integration over the $f_i$ through quadrature.  However, the quadrature technique is valid only for polynomials, so the equilibrium probability distribution function must be expanded in terms of the orthogonal polynomials chosen.

\begin{figure}[h!]
	\centering
	\includegraphics[width=0.4\linewidth]{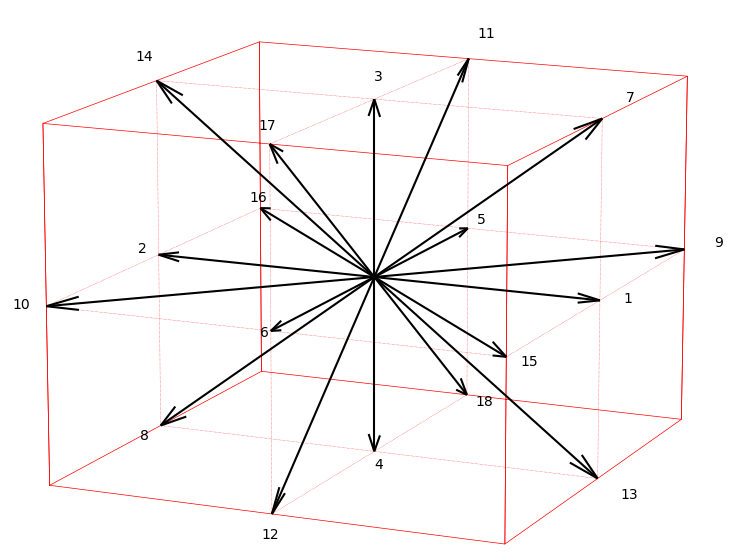}
	\caption[D3Q19 cubic lattice]{Single node of a D3Q19 configured lattice.  The node at the center of the cube is connected to its neighboring nodes by 19 discrete speeds (including 0).}
	\label{fig:d3q19cube}
\end{figure}

Figure \ref{fig:d3q19cube} depicts a single node of a three dimensional lattice with 19 speeds.  Each node contains 19 different probability distribution function values $f_i$, one for each speed at that lattice location.

The principal of dynamic similarity asserts that the fluid equations of motion are the same at all length scales, so it is customary within a LBM formulation to rescale the space and time values to dimensionless quantities with respect to some length $R_{\perp}$ and frequency $\omega_{\perp}$ \cite{2016PhRvA..93a3618B}.
\begin{equation}
	\bar{\bm{x}} = \frac{\bm{x}}{R_{\perp}} ,\ \ \bar{t} = t \omega_{\perp},\ \ 
	\bar{\bm{v}} = \frac{\bm{v}}{R_{\perp} \omega_{\perp}} ,\ \ \bar{\bm{u}} = \frac{\bm{u}}{R_{\perp} \omega_{\perp}} 
\end{equation}
The rescaled quantities are then easily converted to the scale of the domain being modeled when they are recorded.
With both Euclidean space and momentum space discretized, the Lattice Boltzmann Method then simulates the time evolution of a hydrodynamic system with discrete time steps following this process: 
\begin{enumerate}
	\item Initialize the local equilibrium distribution function $f_i^{eq}$ at each node on the grid according to defined initial conditions of the macroscopic variables, and set the new local probability distribution function to be equal to the equilibrium probability distribution function, $f_i^*=f_i^{eq}$. 
	\item \label{list:lbmproc_init_cl} “Stream” the new distribution function $f_i^*$ to the neighboring node connected by the discrete speed, accounting for boundary conditions.
	\item Calculate the macroscopic moments based on the local state at each node, optionally recording them for analysis.  
	\item Calculate the equilibrium probability distribution function $f_i^{eq}$ based on the moments.
	\item Perform the collision calculation locally using the equilibrium probability distribution function $f_i^{eq}$ and the local probability distribution function $f_i$ to find the new probability distribution function:  $f_i^* = f_i + \Delta t C $.
	\item Repeat from step \ref{list:lbmproc_init_cl} for each time step in the simulation.
\end{enumerate}

This technique is advantageous computationally since, though the dynamics of a hydrodynamic system are non-linear (see (\ref{Boltzmann_equation})), the non-linear computations are carried out locally at each node independently and then streamed linearly to the (non-local) neighboring nodes. This allows each node to be computed independently and simultaneously with all the other nodes, providing a high degree of parallelism.  One has only to ensure the simpler streaming step happens serially after the collision calculation for all nodes has been completed.

		\section{Relativistic Lattice Boltzmann}   \label{sec:relBoltz}

The classical version of the LBM has seen wide adoption for the simulation of a variety of hydrodynamic systems.  Many variants have been adopted to improve accuracy or performance or to address other difficulties that can arise in different domains.  Some variants have been proposed to adjust the Lattice Boltzmann Method framework to model relativistic hydrodynamic flows.  These are less abundant in literature than their classical counterpart, but detailed descriptions can be found in the following works: \cite{romatschke2012relativistic}, \cite{weih2020beyond}, \cite{mendoza2010derivation}, \cite{mendoza2010fast}, \cite{blaga2017quadrature},  \cite{li2012lattice}, \cite{gabbana2017towards}, \cite{hupp2011relativistic}, \cite{oettinger2013gaussian} and \cite{bazzanini2021lattice}.  One such variation is proposed by Romatschke, Mendoza and Succi in 2011 \cite{2011PhRvC..84c4903R}.  This Relativistic Lattice Boltzmann Method (RLBM) adaptation is based on the dynamics of the relativistic version of the probability distribution function, $ f=f( x^\mu, p^\nu) $, where $x^\mu$ is the position 4-vector and $p^\nu$ is the momentum 4-vector.  The dynamics of the probability distribution function are described by the relativistic Boltzmann equation with a relativistic analog of the BGK collision term \cite{2002rbet.book.....C}.
\begin{equation} \label{eq:boltzmanneq_rel}
	\left[ p^\mu \nabla_\mu - \Gamma_{\mu\nu}^\lambda p^\mu p^\nu  \partial_{\lambda}^{(p)} \right] f = 
	- \frac{p^\mu u_\mu }{ \tau_R } \left( f - f_J^{eq}  \right)
\end{equation}
For this equation, and for the remainder of this treatment, the units are natural; $ c=k_B=\hbar=1$.  $ \nabla_\mu $ is the covariant derivative, and $\Gamma_{\mu\nu}^\lambda$ is the Christoffel symbols which are given by derivatives of the underlying metric tensor $g_{\mu \nu}$.  For this report we assume Minkowski space-time configuration, so that it is space-time is flat and the second term on the left-hand side is neglected.  The macroscopic fluid velocity is $ u^\nu = \gamma \left( 1, v^i  \right) $, where $ \gamma = \left( 1 - v^2 \right)^{-\frac{1}{2}} $ is the Lorentz factor and $ v^i $ is the fluid 3-velocity.  In this description, Greek indices refer to 4-vector space-time components, and Latin indices are used for 3-vector spatial components.  In the relativistic regime energy, $ E = \sqrt{m^2 + p^2}$, is no longer quadratic in the velocity.  This is reflected in the relativistic form of the equilibrium probability distribution function $f_J^{eq}$ which is the J{\"u}ttner distribution.
\begin{equation} \label{eq:equlibriumdf_rel}
	f_J^{eq}=Z^{-1} e^{ \frac{-p_\mu u^\mu }{T} }
\end{equation}
Here $T$ is the local temperature and $Z$ parameterizes the number of degrees of freedom, which in this work will be taken to be one. 

The macroscopic moments in relativistic hydrodynamics are obtained from the energy-momentum (energy-stress) tensor $T^{\mu \nu} $ which can be obtained by integration of $f$ over the 4 components of momentum space \cite{2011PhRvC..84c4903R}.
\begin{align} \label{eq:energyMomentumIntegral}
	\nonumber T^{\alpha \beta}  \equiv & \int d \chi p^\alpha p^\beta f(t, x, p) \\
	\equiv & \int \frac{d^4p}{\left( 2 \pi \right)^{3}} \delta \left( p^\mu p_\mu - m^2 \right) 2 H \left( p^0 \right) p^\alpha p^\beta f(t, x, p)
\end{align} 
In this definition $m$ is the particle's mass and $H$ is the Heaviside step function.  (Note the notation difference for the relativistic probability distribution function, $f(x^\mu, p^\nu) \rightarrow f(t, x, p) $.)  

Using the J{\"u}ttner form of the equilibrium probability distribution function, integration of (\ref{eq:boltzmanneq_rel}) at equilibrium produces the equilibrium energy-momentum tensor $T_{eq}^{\mu \nu}$ with the familiar relation (see (\ref{stress-energy-tensor-eq}))
\begin{equation} \label{eq:stressEnergyT}
	T_{eq}^{\mu \nu} = \left( \epsilon + P \right) u^\mu u^\nu + P g^{\mu \nu} .
\end{equation} 
Here both energy density  $ \epsilon $ and pressure  $P$ are functions of temperature.  The viscosity coefficient, represented in the collision term, is  $ \eta = \tau_R \frac{\epsilon + P}{5} $, with relaxation time $ \tau_R = 5 \frac{\eta}{s} T^{-1} $, where $s$ is the entropy density \cite{2017arXiv171205815R}.  We apply the canonical, so-called Landau-Lifshitz condition, where $ u_\mu T^{\mu \nu} \equiv \epsilon u^\nu $, to provide a relationship between the energy density and the fluid velocity in the rest frame of reference.

		\section{Boltzmann to Lattice Boltzmann}  \label{section:Boltzmann2LatticeBoltzmann}

This continuous formulation of a relativistic hydrodynamic system is projected onto a discrete lattice for computational modeling following a prescription similar to the classical LBM.  Space and time are discretized in the same manner, but the momentum must be addressed differently.  The relativistic form of the equilibrium distribution function is not quadratic in the exponent, and therefore has a different form than the generator function of the Hermite polynomials; $e^{-\frac{v^2}{T}}\ \rightarrow e^{-\frac{p^\mu u_\mu}{T}}$.  Therefore, the discretization of the velocity space is less straight forward.  To address this, the equilibrium probability distribution function is re-expressed as an expansion about powers of $ \frac{ |\bm{p}| u^0 }{ T_0 \theta } $.
\begin{equation}
	e^{- \frac{p \cdot u}{ T } } = 
	e^{- \frac{ |p| u^0}{ T_0 \theta } } \sum_n 
	\left( \frac{ \bm{p} }{ |\bm{p}| } \right)^n 
	\left( \frac{ |\bm{p}|u^0 }{ T_0 \theta  } \right)^n
	\frac{ \left( \bm{u} \right)^n  }{ \left( u^0  \right)^n n! }
\end{equation}
Here the momentum is expressed as unit vectors, $ \bm{v} = \frac{\bm{p}}{ |\bm{p}| }$, and $ \theta$ is the scaled temperature with respect to a reference temperature $T_0$, $ \theta = \frac{T}{T_0} $.  In this form we can express the relativistic probability distribution function in terms of a (different) set of orthogonal polynomials as:
\begin{equation} \label{eq:rel_pdf_exp}
	f(t,x, p) = e^{ \frac{-p^0}{T_0} } \sum_n P_{ i_1 ... i_n }^{ (n) } \left( \bm{v} \right) a_{ i_1 ... i_n }^{ (n) } \left( t, x, \frac{p^0}{T_0} \right) 
\end{equation}
where $ p^0 = |\bm{p}| $.  The chosen polynomials $ P_{ i_1 ... i_n }^{ (n) } \left( \bm{v} \right) $ are three dimensional polynomials about the unit vector velocity determined by orthogonality with respect to the angular integral $ \int \frac{d\Omega}{4\pi} $.  Their properties are listed in Appendix \ref{appendixA}.  The coefficients in (\ref{eq:rel_pdf_exp}) are determined using orthogonality conditions:
\begin{align} \label{eq:rel_pdf_exp_coef}
	\nonumber a^{(0)} \left(t, x \right) 
	= \frac{1}{\left( \alpha \right)!} \int d\bar{p}\ \bar{p}^\alpha \int \frac{d\Omega}{4\pi} f P^{(0)} \left( \bar{p} \right) , \\
	\nonumber a_i^{(1)} \left(t, x \right) 
	= \frac{3}{\left( \alpha \right)!} \int d\bar{p}\ \bar{p}^\alpha \int \frac{d\Omega}{4\pi} f P_i^{(1)} \left( \bar{p} \right) , \\
	a_{ij}^{(2)} \left(t, x \right) 
	= \frac{15}{2 \left( \alpha \right)!} \int d\bar{p}\ \bar{p}^\alpha \int \frac{d\Omega}{4\pi} f 	P_{ij}^{(2)} \left( \bar{p} \right)
\end{align}
where $\bar{p}=p^0/T_0$ is the scaled momentum magnitude.  The parameter $\alpha$ is chosen to be 3, reflective of the number of dimensions and conveniently mapping the coefficients to elements of the energy momentum tensor.  The angular portion is made discrete using a similar process.  A more complete description of the derivation of the expansion coefficients and the subsequent discretization of the momentum space is described in the work \cite{2011PhRvC..84c4903R}.  

\begin{figure}[h]
	\centering
	\includegraphics[width=0.5\linewidth]{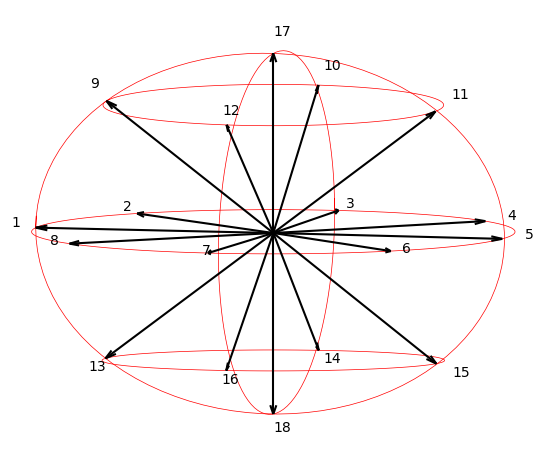}
	\caption[Spherical lattice]{A sketch of a spherical lattice node in an RLBM lattice.  The momentum lattice connecting the nodes is spherical instead of box-like.}
	\label{fig:d3q19sphere}
\end{figure}

Having determined the set of orthogonal polynomials to be used to expand the probability distribution function, a discrete set of momenta can now be determined from their roots using the same quadrature technique described in (\ref{section:stdLBM}).  As the polynomials are orthogonal with respect to the solid angle, the resulting discrete three dimensional momentum lattice is spherical, and is not space filling in general (the connection to neighboring nodes is off-lattice).  This is overcome computationally using a linear interpolation scheme.  An example of a spherical lattice node is shown in figure \ref{fig:d3q19sphere}.

It remains to obtain the macroscopic moments of the fluid system from the probability distribution function by evaluating the energy-momentum tensor.  Using equation (\ref{eq:rel_pdf_exp_coef}) and (\ref{eq:energyMomentumIntegral}) we find:
\begin{equation}
	T^{\mu \nu} = \frac{3 T_0^4}{ \pi^2 } 
	\left(
	\begin{matrix}
		a^{(0)} & a_i^{(1)} \\
		a_j^{(1)} & a_{ij}^{(2)} + \frac{1}{3} \delta^{ij} a^{(0)}
	\end{matrix}.
	\right) 
\end{equation}
The fluid velocity $u^\mu$ is determined by the eigenvalue problem set up by the Landau-Lifshitz condition, $u^\mu T_\mu^\nu = \epsilon u^\nu $, with the energy density as the eigenvalue $\epsilon$.  The temperature is determined from the relation $\epsilon = \frac{3 T^4}{ \pi^2 }$, and pressure $P$ (denoted with a capital $P$ in this section) is found from the equation of state, which is that of an ideal gas in a relativistic system, $P = \frac{\epsilon}{3}$ (see Appendix \ref{finding_eq_st}).  


%
%

\chapter{Part I: Reproduction of a 2D turbulent flow with the RLBM}	\label{stirring_force_project}	

We apply the RLBM hydrodynamic modeler to reproduce the flow of an ideal isotropic fluid system induced to turbulence with a random stirring force, generated in momentum space.  A forcing term is introduced into the relativistic Lattice Boltzmann equation through the discretization of the driving term.  The new driving term is populated with a stirring mechanism driving the fluid in random directions and random (bounded) magnitudes in momentum space at defined frequency ranges to induce an energy cascade, analogous to forced-turbulence studies using the classical LBM for non-relativistic fluids.  The modeler's ability to induce a turbulent response in the flow is evaluated.

	\section{The Driving Term}   \label{section:turbInduceDrivingTerm}
A modification to the RLBM described by \cite{2011PhRvC..84c4903R} is required in order to apply the stirring force to the model.  A turbulent flow is created and sustained by the introduction of a suitably large externally imposed stirring force, zero averaged in the spatial component; $ F^\mu = ( 0, F^i) $.  The force-included relativistic lattice Boltzmann equation in flat space-time is
\begin{equation} \label{eq:boltzmanneq_force_rel}
	\left[ p^\mu \nabla_\mu -  F^\mu \nabla_\mu \right] f = 
	- \frac{p^\mu u_\mu }{ \tau_R } \left( f - f_J^{eq}  \right).
\end{equation}
As with the probability distribution function and the equilibrium probability distribution function, the driving term must be expressed in terms of the orthogonal polynomials 
\begin{equation}
	F^\mu\partial_\mu f\approx e^{-\bar{p}}\sum_{n} {a^{(n)}P^{\left(n\right)}}
\end{equation}
The orthogonality relations again determine the expansion coefficients given in (\ref{eq:rel_pdf_exp_coef}). 

The projected form of the driving term, evaluated in terms of the probability distribution function and the fluid velocity is found to be (see Appendix \ref{appendixB}).
\begin{align} \label{eq:forceTerm}
	\nonumber F_i \partial_p^i f \approx
	\frac{e^{-\bar{p}}}{T_0}  & \left[  
	\frac{3}{12} F_i         \int d\chi\ v^i f
	-           \frac{1}{2}  F^i v^i     \int d\chi\ f \right. \\
	& +  \left. \frac{5}{4}  F_i v_j v_k \int d\chi\ f v^i v^j v^k 
	-           \frac{5}{2}  F_i v_i v_k \int d\chi\ f v^k 
	\right],
\end{align}
where the integration, $ \int d\chi = \int_{0}^{\infty} d\bar{p} \int \frac{d\Omega}{4\pi} $, can be evaluated using the quadrature technique described.  The integrals in (\ref{eq:forceTerm}) are then expressed as expansions with the expansion coefficients in (\ref{eq:rel_pdf_exp_coef}) up to the third order.  The driving term is calculated and applied in the same step of the LBM process as the collision calculation, and is added to the new probability distribution function, $f_i^* = f_i + \Delta t C[f] + \Delta t F^\mu \nabla_\mu f $.  With this modification in place the RLBM is equipped to produce turbulence and to model a turbulent relativistic hydrodynamic system.    

	\section{Model and Methods}

We consider a two-dimensional isotropic hydrodynamic flow of massless particles under a random, zero-averaged stirring force with infinite boundaries.  The RLBM numerical scheme models the system using a spherical D3Q27 momenta configuration connecting $N^2$ lattice nodes, each containing 27 discrete momenta.  The two-dimensional system is represented with a D3Q27 configuration by limiting the number of nodes in one spatial dimension to 1.  The equation of state used in the numerical scheme, that of a three dimensional ideal gas, is applicable for this two-dimensional model given the lattice configuration and the two-dimensional forcing scheme.  The boundaries of the lattice are periodic and the initial state sets the probability distribution of the momenta at each point as flat; i.e. equivalent at each node on the lattice. Simulations were conducted with lattice size $N$ set to $128, 256, 512,$ and $1024$ and the lattice spacing $\alpha$ set to $0.32, 0.16, 0.08$ and $0.04$ lattice units respectively, so that lattice configurations model systems with constant volume.  The viscosity, represented as the ratio of viscosity to entropy $\frac{\eta}{s}$, is set to $0.01$ in the collision term.    

Simulations begin initially at rest, and turbulence is introduced into the system as a response to a stirring force applied in Fourier space in zero-averaged random spectral directions and at a defined range of wave numbers.  A filter in Fourier space restricts the driving wave number range from a chosen wave number magnitude $k_f$ to the largest wave number magnitude permitted by the discrete Fourier lattice.  Therefore, the driving range is defined as $ k_f \leq k < k_{max} $, leaving the inertial range as $0 < k < k_f $.  Most tests were performed with $k_f$ set to $6$ in lattice units, and the maximum wave number $k_{max}$ varying with the lattice spacing.  At each node in the spectral lattice within the driving range, a driving term is applied at a random magnitude up to a maximum $|F|$, and in a random spectral direction within the plane so that it averages to zero throughout the lattice.  The maximum forcing magnitude is chosen to produce an adequate fluid velocity in the inertial range to induce a turbulent flow.  The forcing spectral lattice is then transformed to configuration space and applied to the relativistic lattice Boltzmann equation as a part of the forcing term.  Each simulation is run in $0.05$ time unit increments for $50.0, 100.0, 200.0$ or $400.0$ lattice time units.  Macroscopic variable calculation is conducted at each time step and recorded every few time steps.

\begin{figure}
	\centering
	\includegraphics[width=0.75\linewidth]{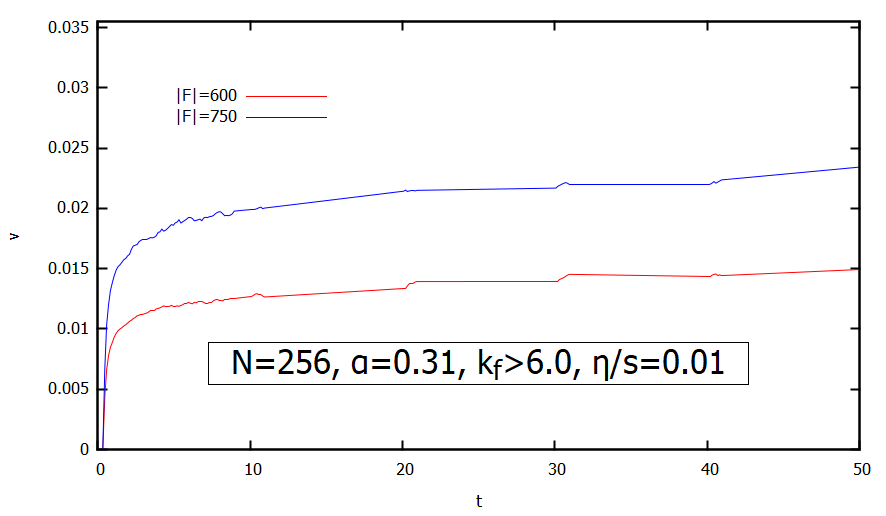}
	\caption[Average fluid velocity]{Filtered average fluid velocity throughout the execution of a simulation at different maximum driving force magnitudes.  The simulations were executed with $256^2$ lattice nodes and a lattice spacing of $0.31$.  The average velocities are stable after about 32 simulated time units.  Figure reproduced from \cite{watson2021two}.}
	\label{fig:velocity-0256}
\end{figure}

\begin{figure}
	\centering
	\includegraphics[width=0.75\linewidth]{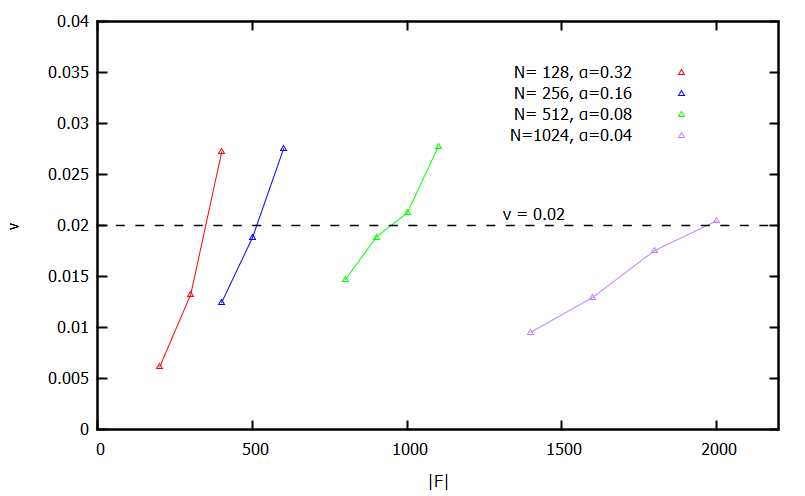}
	\caption[Determination of a constant velocity]{A constant filtered average velocity is determined across multiple lattice configurations empirically using a plot as shown.  A variation of the lattice spacing in the lattice configuration changes the number of discrete wave numbers, changing the size of the driving filter and affecting the average velocity of the simulation.  A constant velocity across multiple lattice configurations is determined by finding a common constant velocity simulated with varying maximum force magnitudes.  Figure reproduced from \cite{watson2021two}.}
	\label{fig:find-avg-velocity}
\end{figure}

The fluid velocities recorded are filtered to eliminate the contribution of the driven spectral range.  This provides a better analysis of the effects of the propagation of energy into the inertial spectral range, but note that recorded fluid velocities represent only a portion of full average fluid velocities.  The velocities obtained for each test varied based on the size of the stirring force filter and the maximum magnitude applied.  They were observed to be stable and consistent after a sufficient amount of time (Fig \ref{fig:velocity-0256}).  The achievement of stable velocities allows the filtered average velocity to be effectively tuned as a parameter so as to maintain a consistent value between simulations with different configurations.  It was tuned by selecting an appropriate stirring force maximum magnitude $|F|$.  This was done empirically by comparing the average fluid velocity of simulations with different forcing configurations for a given lattice and matching it with those of other lattices.  Figure \ref{fig:find-avg-velocity} shows an example of a line of constant velocity between four different lattice configurations with different forcing filters.    

 \hfill  

	\section{Results}
For modest driving force magnitudes, the numerical model produced turbulence inducing fluid velocities of $\sim 0.01$ (all values in this section are in lattice units unless otherwise specified) in the inertial range, but yielded no turbulent signal when the driving magnitude was too small.  In an inviscid system even small velocities can induce turbulence.  Indeed, in the absence of viscosity, the vortices can grow unbounded so that, in many computational fluid dynamic codes modeling an Euler fluid a viscosity is induced numerically to ensure stability.  But the model simulations suggest the existence of a minimum velocity to show energy propagation across the spectrum in a system with non-zero viscosity.  For a lattice spacing of $0.01$, a minimum viscosity of $0.005$ was needed to keep the evaluation of the probability distribution function numerically stable, but a smaller viscosity did not affect the stability of models with smaller lattice spacings.  The maximum viscosity was found to be $0.18$ at the same lattice spacing, also determined by stability.  Some other instabilities were observed as a result of excessively high fluid velocities present in the driven range, which are required to induce energy propagation to the inertial range.  

\begin{figure*}
	\centering
	\includegraphics[width=1.0\linewidth]{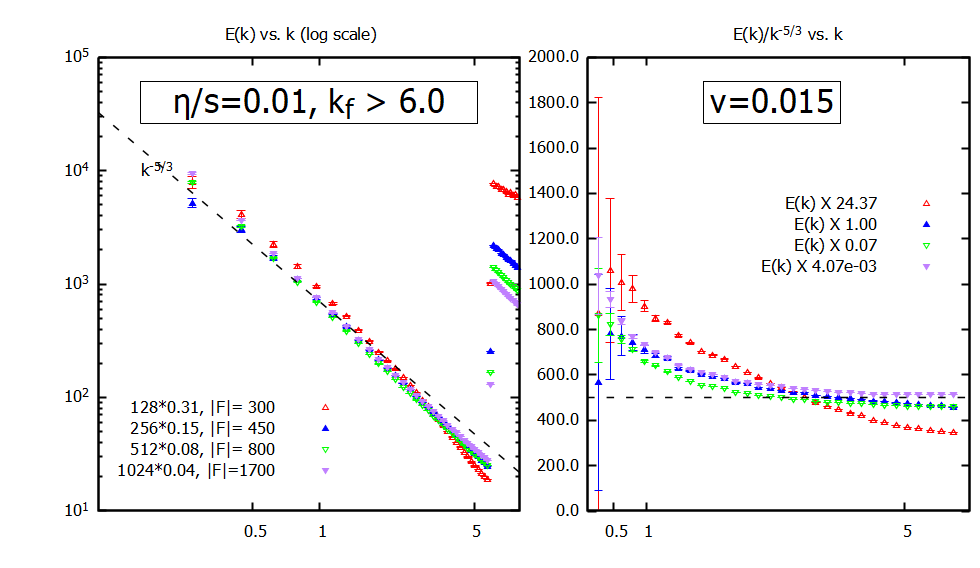}
	\caption[Energy spectrum for $k_f > 6$]{The energy spectrum for simulations with lattice configurations of $N\alpha = (128)(0.32), (256)(0.16), (512)(0.08)$, and  $(1024)(0.04)$, keeping a constant volume.  The energy for each lattice configuration is scaled to a similar magnitude so the curves can be easily compared.  The left panel shows the slope of the energy spectrum at log scale as compared to the expected power law of $-5/3$ in the inertial range.  Part of the driven range is visible at $k>6.0$, marked by a discontinuity in the slope to a much higher energy value.  The right panel is a "close up" of the same divided by the expected slope.  Close conformity to the $-5/3$ slope is noted, with deviations at larger scales, and for configurations with larger lattice spacing.    Figure reproduced from \cite{watson2021two}.}
	\label{fig:hiband-filter-const-vol}
\end{figure*}

With adequate fluid velocity the RLBM numerical scheme shows evidence of the spectral energy propagation caused by turbulence, reflecting the slope of the energy in k-space described by Kolmogorov.  Figure [\ref{fig:hiband-filter-const-vol}] shows the power spectrum for systems modeled with different lattice configurations maintaining a constant volume.  Lattice configurations with $N$ set to $128, 256, 512$, and $1024$ are shown with lattice spacings of $0.32, 0.16, 0.08$, and $0.04$ respectively.  The driving filter is applied at $k_f = 6.0$, and at a maximum forcing magnitude determined for each to produce a constant velocity of $0.015$.   The left panel is the energy spectrum at log scale comparing the slope of the power spectrum against Kolmogorov’s predicted slope of $-5/3$.  The right panel is a “zoomed in” view of the same spectrum divided by the Kolmogorov slope in order to highlight the conformity.  The energies of the various configurations are scaled independently so that they align on the plot for comparison.

 \hfill

A power spectrum cascade was observed with a reasonable reproduction of Kolmogorov's power law in the inertial range for mid and large wave numbers.  Deviations at large scales ($ \sim N\alpha $) is noted, and is presumed to be a product of the limited discrete spectrum causing a “pileup” at the largest discrete scales.  The effect is eased by changing the lattice spacing to create more small $k$ modes below the driving range, giving the energy more modes to which it can propagate.  Because of the larger errors, the results at that scale of the system are not considered in the conclusions.  The simulation configured with $N = 128$ and $\alpha=0.32$ shows a significant pileup at the large scales and lower conformity to the $-5/3$ power law at small scales.  The smaller number of nodes requires a larger lattice spacing to maintain the same  volume, which maximizes the distortion of a discrete lattice and impedes its ability to accurately model a continuous system.  

\begin{figure*}
	\hfill
	\includegraphics[width=1.0\linewidth]{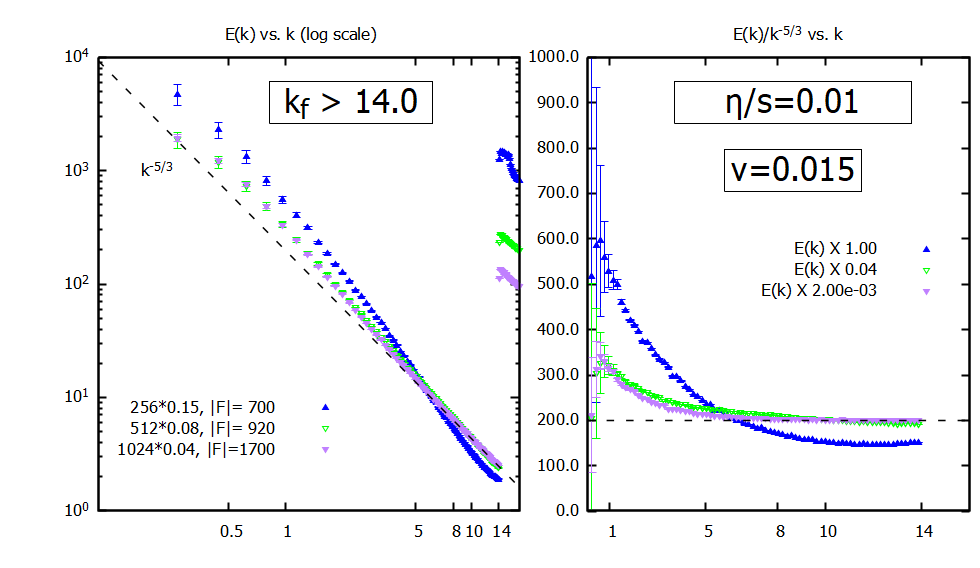}
	\caption[Energy spectrum for $k_f > 14$]{The energy spectrum of the inertial range $0 < k < 14.0$ for simulations with lattice configurations of $N\alpha = (256)(0.16), (512)(0.08)$, and $(1024)(0.04)$, scaled for comparison.  The left panel shows the slope of the energy spectrum at log scale as compared to the expected power law of $-5/3$, and the right panel is a "close up" of the same divided by the expected slope.  A larger inertial range shows improved conformity.    Figure reproduced from \cite{watson2021two}.}
	\label{fig:hiband-kf14-constVol}
\end{figure*}

As the inertial range of the spectrum expands the power spectrum forms a better resemblance to the $-5/3$ power law and the distortion at the large scales is eased, implying a closer conformity at the continuum limit.  That is, as the number and range of discrete wave numbers in the inertial range grows the slope of the power spectrum approaches a closer resemblance to what is expected in a continuous system.  Figure \ref{fig:hiband-kf14-constVol} compares the same lattice configurations as in figure \ref{fig:hiband-filter-const-vol}, but driven with a forcing configuration leaving the inertial range from $k = 0$ to $k = 14$.  This leaves the $256^2$ node configuration simulation driving at only a third of the accessible wave numbers which disrupted the spectral slope.  But the spectra of the larger lattice configurations show a much better conformity to the power law in a larger portion of the inertial range.  Simulations also showed that the larger the inertial range, the more kinetic energy injected into the system is required to achieve the energy propagation.  Indeed, a fluid velocity of $\approx 0.25$ was stable for a $N\alpha = (1024)(0.04)$ configuration with a forcing configuration of $k_f=60$ and $|F|=6000$, where the driven range is only 30\% of the available spectrum.  

\begin{figure*}
	\centering
	\includegraphics[width=1.0\linewidth]{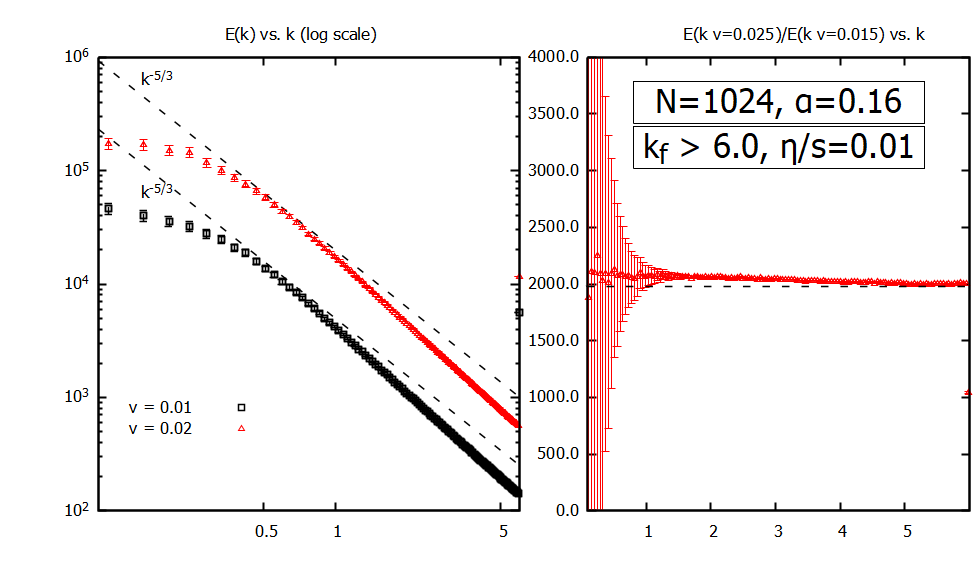}
	\caption[Energy spectrum for varying velocities]{The energy spectrum of a simulation with lattice configuration $N\alpha = (1024)(0.16)$ with large and small filtered average velocity.  The left panel compares the energy spectrum at log scale with the expected slope of $-5/3$.  The right panel is the energy spectrum of the larger velocity simulation divided by the smaller velocity simulation.  The magnitude of the fluid velocity does not have an effect on the inverse energy cascade, provided the velocity is sufficient to produce turbulence.    Figure reproduced from \cite{watson2021two}.}
	\label{fig:hiband-hilo-multiN}
\end{figure*}

The energy spectrum of a larger velocity flow is compared to a smaller velocity flow for a $ N\alpha = (1024)(0.16)$ lattice configuration in figure [\ref{fig:hiband-hilo-multiN}].  The left panel is the energy spectrum at log scale, and the right panel is the energy spectrum of the larger velocity system divided by the smaller one which is rendered as a zero-slope line.  No discernible differences in the energy spectrum’s slope due to velocity is detected, providing evidence that the numerical model's energy propagation is not dependent on the average velocity, provided it is significant enough induce energy propagation.

\begin{figure*}
	\centering
	\includegraphics[width=0.75\linewidth]{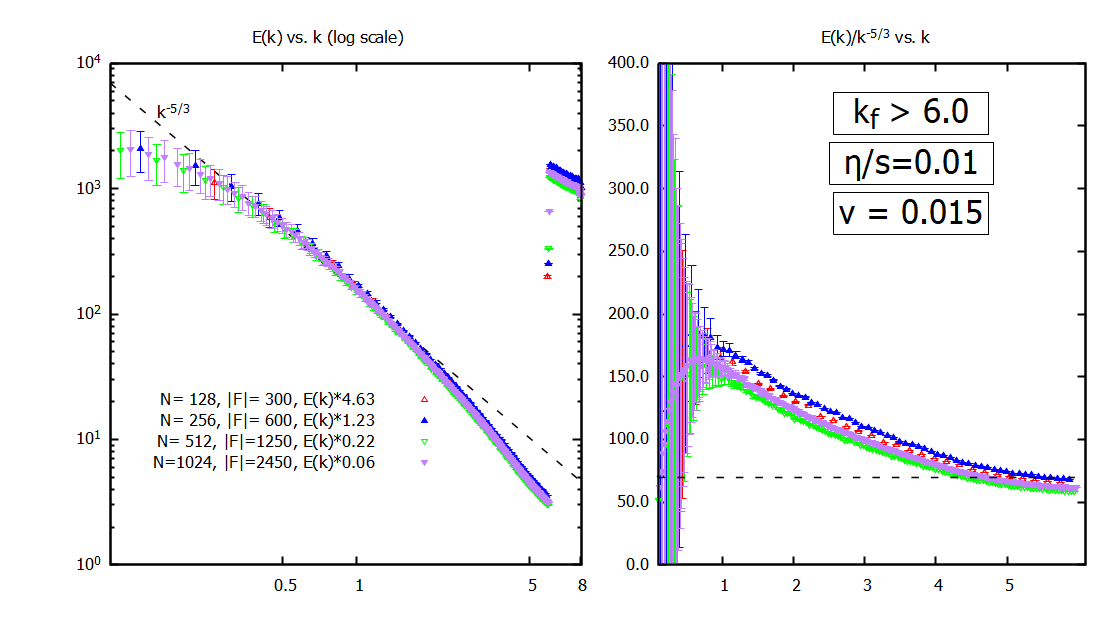}
	\caption[Energy spectrum for varying volume]{The energy spectrum of lattice configurations $N\alpha = (128)(0.16), (256)(0.16), (512)(0.16)$, and $(1024)(0.16)$ (increasing volume), with a constant filtered average fluid velocity of $0.015$.  The left panel compares the energy spectrum to the expected slope of $-5/3$ at log scale, and the right is the energy spectrum divided by the expected slope.  The energy spectrum in each panel is scaled for comfortable comparison.  The similarity of the slope of each curve suggests the energy propagation is not sensitive to volume when the volume is controlled by the number of lattice nodes.    Figure reproduced from \cite{watson2021two}.}
	\label{fig:hiband-filter-incr-vol}
\end{figure*}

\begin{figure*}
	\begin{minipage}{0.48\textwidth}
		\centering
		\includegraphics[width=1.0\linewidth]{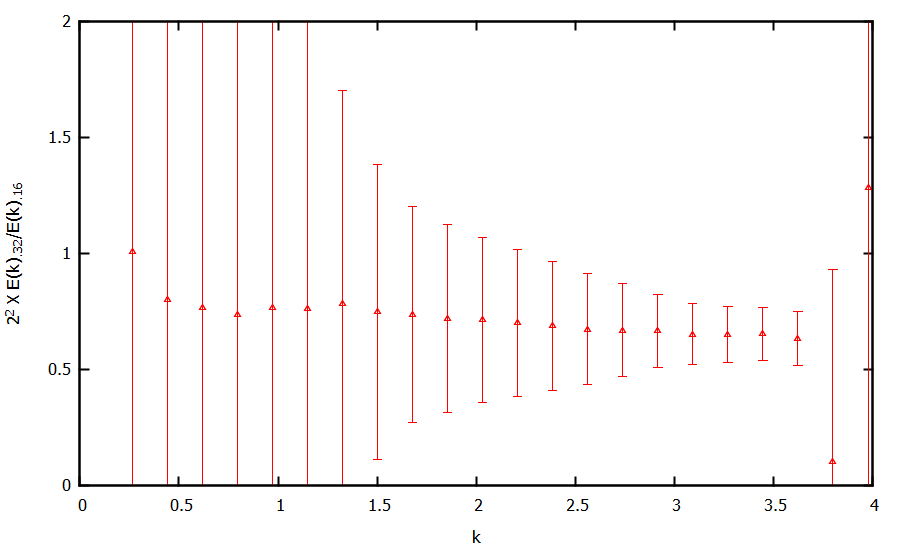}
		\label{fig:ratio-latspc-diffvol}
	\end{minipage}
	\begin{minipage}{0.48\textwidth}
		\centering
		\includegraphics[width=1.0\linewidth]{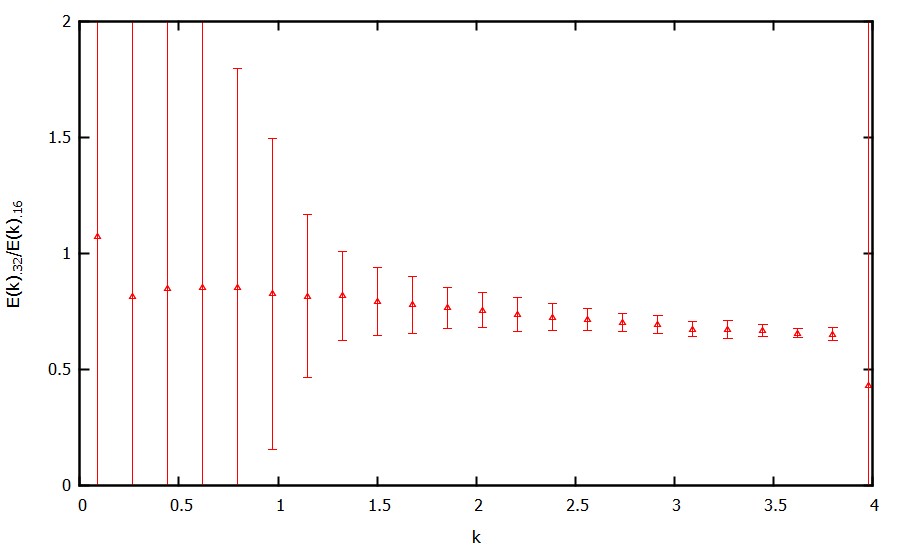}
		\label{fig:ratio-latspc-samevol}
	\end{minipage}
	\caption[Energy spectrum lattice spacing ratio]{Left is the ratio of the power spectrum $E(k)$ for simulations with lattice spacing $0.32$ and $0.16$ and a constant $N$.  The ratio at each $k$ trends to constant, suggesting the simulated results are not affected by lattice spacing alone. The $\sim 3/4$ ratio is attributed to an additional mode available with a tighter lattice. The lattice with the smaller volume is scaled with a factor of $2^2$ to compensate for the larger energy capacity of the larger volume system.  Right is the same power spectrum ratio for simulations with lattice configurations $N\alpha = (128)(0.032)$ and $(128)(0.016)$.  The ratio also tends to be constant, suggesting the simulated results are not affected by the non-physical lattice parameters $N$ and $\alpha$.  Figure reproduced from \cite{watson2021two}.}
	\label{fig:ratio-latspc-diffsamevol}
\end{figure*}

Figure [\ref{fig:hiband-filter-incr-vol}] shows the energy spectrum for systems of increasing size, modeled by an increasing number of lattice nodes with constant lattice spacing.  The conformity to the expected slope mimics the simulations performed with a constant volume implying the model's propagation of energy is not dependent on the number of nodes, a non-physical parameter.  Figure (\ref{fig:ratio-latspc-diffsamevol}) (left) shows the scaled ratio of $E(k)$ observed for a simulation with a lattice spacing of $0.32$ to $E(k)$ for a simulation with a lattice spacing of $0.16$.  The absence of a slope for the ratio suggests the power spectrum is not sensitive to lattice spacing alone.  A similar comparison (right) of the energy spectrum for systems with a constant volume yields a similar ratio.  Finally, the energy propagation was found to be independent of the configuration of the stirring force (the choice of $k_f$, effective $k_{max}$, and $|F|$) provided it is large enough to induce a sufficient velocity capable of producing turbulence.

	\section{Conclusions}
	
The energy spectrum of the modeled flow demonstrates agreement with Kolmogorov’s predicted slope of $-5/3$ in the inertial range indicating an inverse cascade of energy expected for a turbulent flow.  The error is large at large scales and shows evidence of a pileup, particularly for lattice configurations with a smaller number of scales in the infrared range.  But the slope is reasonably reproduced in the inertial range above the large scales up to the driven scales, implying the model is a reasonable reproduction of turbulent two-dimensional flows of a relativistic fluid.  The model is shown to be insensitive to non-physical parameters including lattice size except with respect to the effects of a large lattice spacing.

%
%
		
\chapter{Part II: Reproduction of turbulence in graphene}    \label{obstacle_project}

Graphene is a single atom thick sheet of graphite made up of carbon atoms in an honeycomb lattice with interesting conductive properties.  The linear dispersion relation of its band structure transports the electric signal through massless quasi-particles, and the flow of the particles can be analyzed through a relativistic hydrodynamic mechanism.  A sample of graphene will contain impurities embedded within the lattice structure or within the supporting substrate that affect the current based on their number density, size, placement and electric properties.  In a hydrodynamic flow, if the average fluid speed around these obstructions is large enough compared to the viscous damping effect, a turbulent flow is possible.  

\begin{figure*}[ht] 
	\centering
	\includegraphics[width=2.25in, height=1.85in]{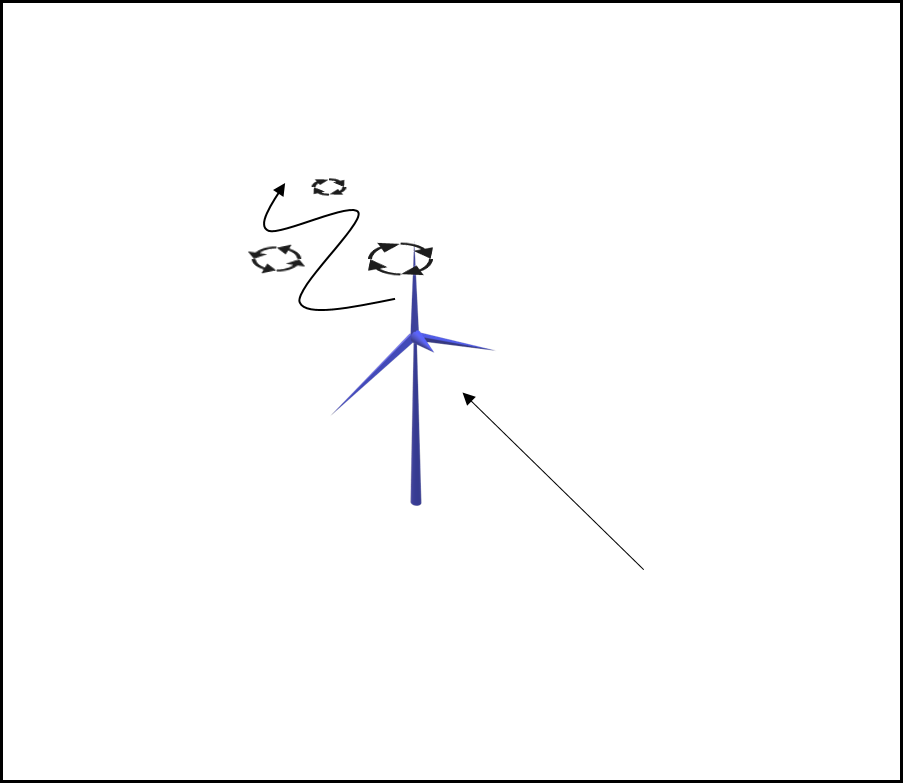}
	\includegraphics[width=2.25in, height=1.85in]{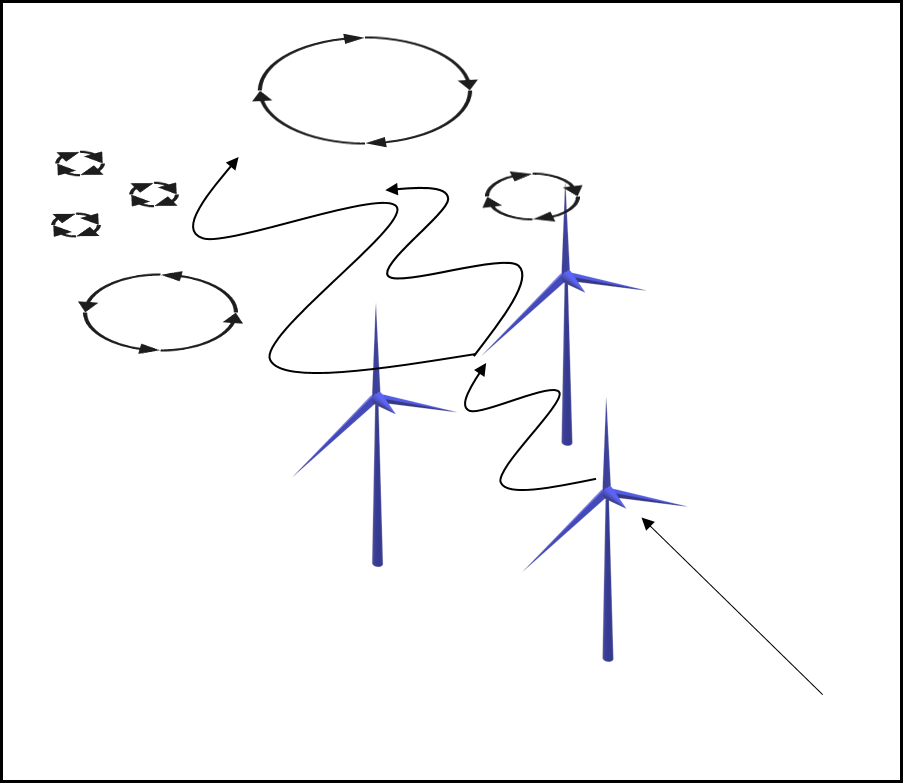}
	\caption[Airflow in a wind farm]{The left shows a vortex street effect a single wind turbine can have on the air flow in its wake when the wind velocity is large enough.  The right sketches a turbulent wind flow caused by additional wind turbines positioned in the wake of another.  The vortex street can turn into a turbulent flow.}
	\label{figure:wind-farm-arrangement}
\end{figure*}

In this section we note a dependency of turbulence in the Dirac fluid in graphene on the position of an obstruction relative to other impurities present in the sample.  The effect is similar to the susceptibility of a wind farm to turbulence based on the placement of the turbines.  An engineer must take into account the effect the wake a turbine has on downwind turbines when determining their placement.  At high wind velocities a single turbine can create a vortex street, or regularly sized and spaced vortices, within the flow of its wake (fig. \ref{figure:wind-farm-arrangement} left).  If another turbine is placed relatively close to its upstream neighbor near its wake, the combined effect may produce a turbulent flow (fig. \ref{figure:wind-farm-arrangement} right) which will affect the efficiency of the farm \cite{stevens2017flow} \cite{ahsbahs2020wind}. 

Though charge neutral graphene is an exceptionally conductive material and the charged quasi-particles are able to obtain a large velocity compared to the viscosity, it is believed that it is unlikely to be capable of producing a turbulent flow under normal circumstances \cite{Lucas_2018}.  However, a recent study by Mendoza, Herrmann, and Succi \cite{PhysRevLett.106.156601} modeled a 5 $\mu m$ impurity obstructing a quasi-particle flow of velocity $10^5$ $m/s$ using a two-dimensional relativistic hydrodynamic numerical model \cite{gabbana2018numerical}.  They calculate the Reynolds number to be on the order of $100$ and proposed that, under circumstances such as these, graphene could potentially produce a flow that is in the preturbulent range.  The present report employs a similar hydrodynamic modeler based on the lattice Boltzmann method to reproduce the flow of a charged Dirac fluid in an experimentally realistic sample of graphene, incorporating impurities in the substrate or embedded within the honeycomb lattice.  Though the Reynolds number calculation in this work differs from what was obtained in \cite{PhysRevLett.106.156601}, we find a signal of a possible preturbulent flow under specific conditions related to the placement of the impurities.   

We employ a hydrodynamic modeler similar to \cite{PhysRevLett.106.156601} and described in \ref{section:Boltzmann2LatticeBoltzmann} to reproduce the flow of a charged Dirac fluid in an experimentally realistic sample of graphene, incorporating impurities embedded within the honeycomb lattice or in the substrate.  Using experimentally realistic parameters, though the estimate for the Reynolds number in this work is smaller, we find a signal of possible preturbulent flow, corroborating the results of \cite{PhysRevLett.106.156601}.   What is new in this work is that we also consider \emph{multiple} impurities in a single sample and investigate if a relativistic flow around these can give rise to turbulent (as opposed to preturbulent) flow signals.

	\section{Model and Methods}

The Reynolds number provides the most commonly accepted metric to predict the presence of turbulence.  Classically the Reynolds number is determined, as stated, using the kinematic viscosity $\nu$ defined as the ratio of the dynamic viscosity $\eta$ to the fluid's mass density $\rho$; $\nu = \frac{\eta}{\rho}$.  For fluid systems consisting of massless particles, formulations of the Reynolds number commonly replace the mass density with the entropy density $s$.  Therefore, the expression for the dimensionless viscosity is $\frac{\eta}{s}$, and the Reynolds number is expressed as $Re = \frac{UL}{(c^2 / T) (\eta / s)}$, where $\frac{c^2}{T}$ balances the units.  The Reynolds number found in \cite{PhysRevLett.106.156601} is determined with this formulation.  Alternatively, one is able to retain the classical expression of kinematic viscosity for a fluid of massless particles by defining the mass density in terms of the number density of the quasi-particles and their ``effective'' mass; $\rho = n m_e$.  The number density is determined experimentally using techniques such as the measurement of the Hall resistance, while the effective electron mass is determined through methods such as the Shubnikov-De Haas effect.  The resulting Reynolds number is a ratio incorporating the velocity of a volume of massless particles with respect to inertial mass to the kinematic viscosity defined in terms of the effective mass with respect to the particles' electric properties.  This more classical form of the Reynolds number is more readily compared to that of traditional fluids.  

Turbulence can also be identified through its effect on fluctuations in the current density.  A turbulent flow creates vortices within the fluid creating multiple traveling modes in Fourier space producing a broadband signal.  Therefore, if the spectrum in frequency space and wavenumber space are both broadband, the flow can be considered turbulent.

To explore the possibility of a turbulent flow of the charged quasi-particles in a sample of graphene under realistic conditions we simulate the hydrodynamic equations of motion using the described adaptation of the relativistic lattice Boltzmann method defined by Romatschke, Mendoza and Succi \cite{2011PhRvC..84c4903R} (see \ref{section:Boltzmann2LatticeBoltzmann}).  The RLBM model reproduces a two-dimensional quasi-particle flow in a $40\mu m $ by $10\mu m$ sample of graphene with one or two rigid impurities obstructing the flow.  The sample is simulated with a $1024 \times 256$ node lattice with a spacing of $0.038 \mu m$.  Initial tests incorporate a single circular impurity with a diameter $D=0.5 \mu m$ placed within the sample in a region near the inflow boundary, referred to as the ``obstruction region''.  Subsequent tests are conducted with a second impurity placed at controlled distances from the first within the same region.  At normal temperatures the impurities within a graphene sample are believed to be charged, creating charge puddles in the surrounding region that affect the electric flow.  The impurities are largely sourced from the substrate \cite{PhysRevLett.98.076602} \cite{Adam18392}, but can also be embedded within the sample itself.  The size of the impurities can vary greatly depending on the foreign material, but they typically stay below approximately $0.5 \mu m$ \cite{Lucas_2018}.  The size and placement of the impurities are difficult to control in an experimental setting, but the model seeks to simulate the effects of one or two quasi-isolated impurities on the current density in ideal but realistic conditions in order to determine if the detection of a turbulent signal is possible.  Therefore the diameter of the obstacle is chosen to maximize its turbulence producing potential while maintaining a realistic size.  The velocity of the charged flow in the Dirac liquid can be relatively high owing to a large effective electric coupling constant.  Flow speeds on the order of $10^2$ $m/s$ are common \cite{10.1038/nnano.2008.268} \cite{doi:10.1063/1.3483130}, but the flow can approach velocities as large as 10\% of the Fermi velocity, $v_F\approx 1.1 \times 10^6 m/s$ \cite{Lucas_2018}.  In order to maximize the possibility of a turbulent signal, the model introduces the largest realizable fluid velocity of $10^5$ $m/s$ ($0.1$ $v_F$) into the sample at the same magnitude along the inflow border.  The borders that are perpendicular to the inflow border use periodic boundary conditions, effectively simulating an infinitely wide sample with multiple, regularly placed obstacles, but at a distance where the wakes created by the obstacles cannot affect each other.  Each lattice node in a region occupied by an impurity implements bounce-back boundary conditions.

The Reynolds number is determined for each test to predict the presence of turbulence using the kinematic viscosity based on the number density as described.  The kinematic viscosity in graphene is found to be $0.000132$ $\frac{\hbar c^2}{eV}$ in natural units ($c = k_B = \hbar = 1$) \cite{PhysRevLett.103.025301}, readily obtained from the value of the diffusion constant for momentum $\eta = 2.633\times 10^3 \frac{eV^2}{c^2 \hbar}$ found in \cite{PhysRevLett.103.025301}, the number density $n = 38.93 \frac{eV^2}{\hbar^2 c^2}$ determined in \cite{PhysRevLett.99.226803}, and the effective mass of a charged Dirac quasi-particle participating in the electric flow, usually given in terms of the mass of an electron $m_e$, where $m_e = 5.11 \times 10^5 \frac{eV}{c^2}$.  The effective electron mass is taken here to be $1.0$ $m_e$.  The diameter of the embedded impurity is the most appropriate choice for the system's characteristic length for the Reynolds number formulation; $U = D$.  

The state data for each node in the lattice is collected throughout the simulation and the local macroscopic moments are determined and recorded.  The current density $\mathbit{j}$ is calculated along the lattice nodes at the outflow border and the frequency of the fluctuations is determined in Fourier space against time.  The spatial fluctuation of the current density is determined along the lattice nodes in the region of the sample down stream from the current inflow referred to as the ``current density sampling region''.  The spectrum of the current density fluctuations in frequency space and in k-space are recorded and plotted for qualitative inspection to look for evidence of mode generation and migration, indicative of a turbulent flow.  

	\section{Results}

		\subsection{Single Impurity}
\begin{figure*}[h!]   
	\includegraphics[width=0.95\linewidth]{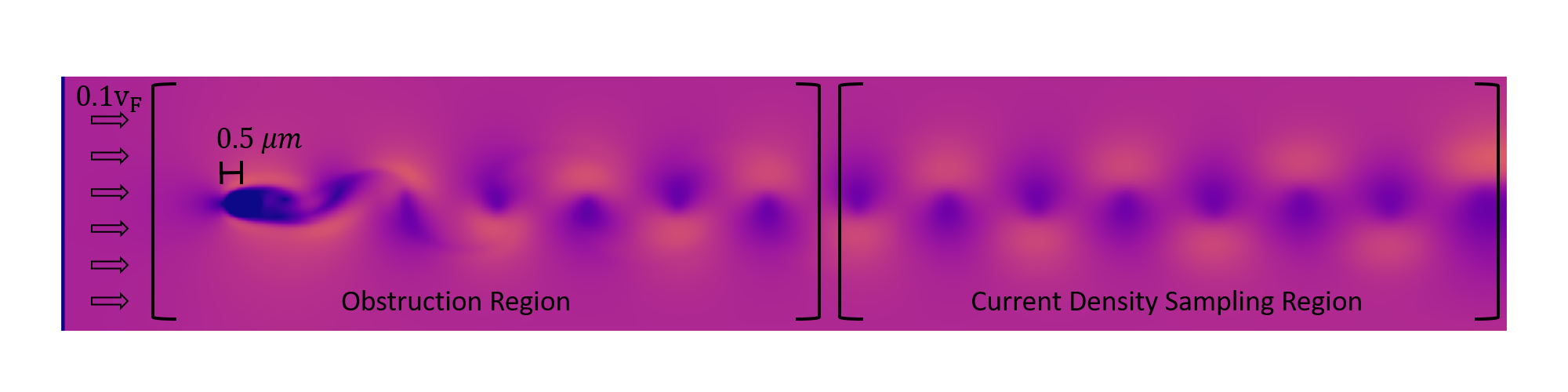}
	\caption[Charged flow about a single impurity]{A velocity magnitude heatmap depicting the flow of charge carrying quasi-particles on a simulated sample of graphene at $0.1 v_F$ around an impurity of diameter $0.5 \mu m$ in the obstruction region.  A large velocity around the obstacle creates a regular pattern of vortices in the wake affecting current density fluctuations detected in the sampling region.  The von K{\'a}rm{\'a}n vortex street flow pattern is too regular to be considered turbulent, producing only a single mode in wave number space.}
	\label{single_heat}
\end{figure*}

\begin{figure}[!h]  
	\centering
	\includegraphics[width=0.8\linewidth]{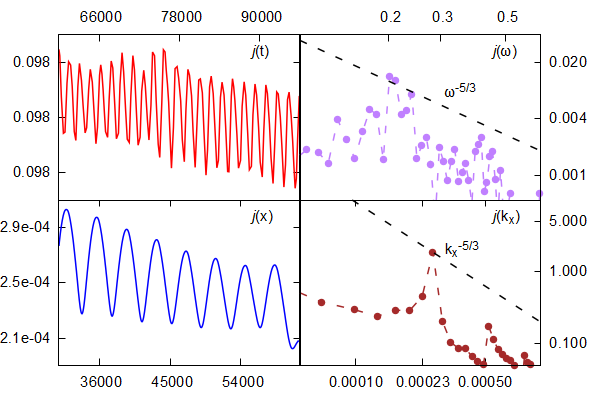}
	\caption[Current density from a single impurity]{The effects of a single impurity of diameter $0.5 \mu m$ on the current density with respect to time $j(t)$ (top left) and space $j(x)$ (bottom left).  The frequency spectrum of $j(\omega)$ (top right in log scale) shows one or two prominent modes that are beginning to migrate.  One prominent mode is seen in the wave number spectrum of $j(k_x)$ (bottom right in log scale) indicating the chaotic flow does not show signs of turbulence.  The current density $j$ is in units of the Fermi velocity $v_F$, and $t, x, \omega,$ and $k_x$ are in modified natural units (see Appendix \ref{appendixA}). }
	\label{single_cur_dens}
\end{figure}

A simulation of a Dirac fluid flow in a sample of graphene containing a single impurity of diameter $0.5 \mu m$ shows a breaking of longitudinal symmetry that develops into vortices in the wake, and forms a flow pattern known as a von K{\'a}rm{\'a}n vortex street.  The vortices form on both sides of the obstacle's wake at a roughly consistent size and placement, alternating on either side of the wake, and shedding at regular intervals (fig. \ref{single_heat}).  The vortex creation, coherence	 and shedding produces temporal fluctuations in the current density that are detected at the outflow border and spatial fluctuations found throughout the sampling region (fig. \ref{single_cur_dens}).  The fluctuations produce one or perhaps two prominent modes in the frequency spectrum, but they appear to broaden to create a slope of about $\omega^{-5/3}$ in small portions of the spectrum. There is, however, a single prominent mode in the wave number spectrum indicating the flow around the obstacle does not produce a turbulent signal in the current density.  The Reynolds number for this system is readily calculated in terms of modified natural units based on the Fermi velocity (see Appendix \ref{appendixA}) using the diameter of the impurity as the characteristic length ($L = D = 0.5\mu m = 3562.5 \frac{\hbar v_F}{eV}$), the average flow speed ($U=0.1 v_F$), and the number-density-dependent kinematic viscosity ($\nu = 11.8 \frac{\hbar v_F^2}{eV}$).  It is found to be 
\begin{equation} \label{reynolds_number_ndkv}
	Re = \frac{ L U }{ \frac{\eta}{\rho} } =  \frac{ \left( 3562.5 \frac{\hbar v_F}{eV} \right) \left( 0.1 v_F \right) }{ \left(11.8 \frac{\hbar v_F^2}{eV} \right) } = 6.37.
\end{equation}
It is near the range for preturbulence, but a von K{\'a}rm{\'a}n vortex street in the flow is a preturbulence phenomenon.  Therefore we conclude the single impurity in this system creates a preturbulent flow, but not turbulence. 

		\subsection{Multiple Impurities}

\begin{figure*}[ht] 
	\includegraphics[width=0.95\linewidth]{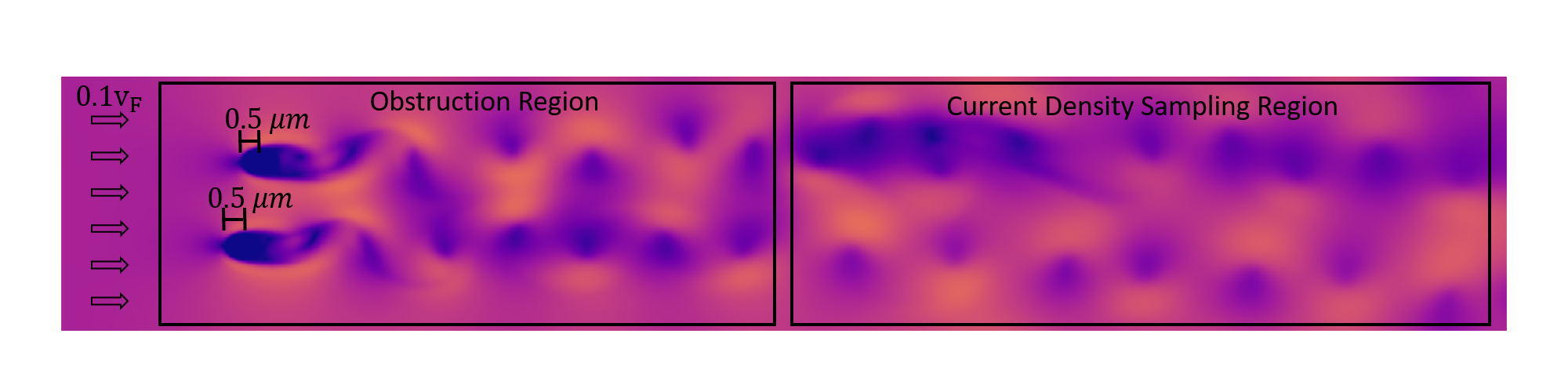}
	\caption[Charged flow about two impurities]{A velocity magnitude heatmap of a quasi-particle flow around two impurities of size $0.5 \mu m$, showing an irregular pattern of vortices in the wake of the two obstacles suggesting turbulence may be present in the flow.  Vortex creation and shedding is evident as in the same system modeling a single impurity, but the irregularity of the coherence of the vortex structures cause a much less regular pattern in the fluctuations in the current density.}
	\label{2obsbest_obs_heat} 
\end{figure*}

\begin{figure}[!h]  
	\centering
	\includegraphics[width=0.8\linewidth]{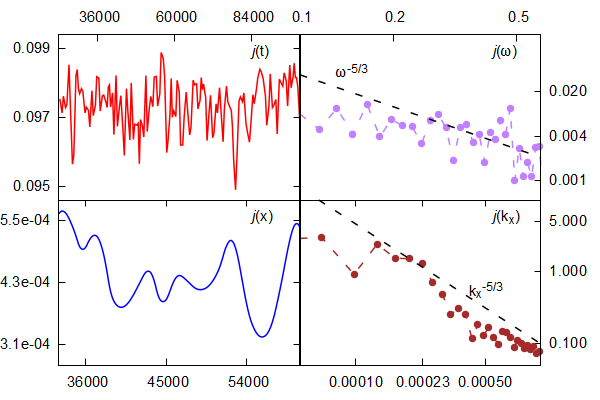}
	\caption[Current density from two impurities]{The current density of a quasi-particle flow around two impurities of size $0.5 \mu m$ with respect to time $j(t)$ (upper left), with respect to space along the axis parallel to the flow $j(x)$ (lower left), with respect to frequency $j(\omega)$ (upper right in log scale), and with respect to wave number $j(k_x)$ (lower right in log scale).  The current density in frequency space, $j(\omega)$, shows the formation of multiple modes propagating from a single mode creating a rough conformity to a $-5/3$ slope in log scales.  The current density in k-space, $j(k_x)$, is broadband with a close adherence to a $|k|^{5/3}$ slope expected for migrating modes in a non-linear system.  The broadband spectra in frequency space and in k-space hint at a potential turbulent signal.  The current density $j$ is in units of the Fermi velocity $v_F$, and $t, x, \omega,$ and $k_x$ are in modified natural units (see Appendix \ref{appendixA}).
	}
	\label{2obsbest_cur_dens}
\end{figure}

A second obstacle is introduced in a subsequent model with the same impurity size and lattice spacing as in the initial test, $0.5 \mu m$ and  $0.038 \mu m$ respectively.  The new obstacle is placed next to, but slightly offset behind the first with respect to the flow in the obstruction region at a separation distance of about $3.38 \mu m$ (fig. \ref{2obsbest_obs_heat}).  The Reynolds number for this system, found to be $12.74$, is similar to the ratio determined for the initial, single-obstacle test and falls to the edge of the lower boundary of what can be considered preturbulent.  However, we see a less regular vortex pattern and the von K{\'a}rm{\'a}n vortex street flow pattern is no longer present.  The fluctuations in the current density show emerging modes in both frequency space and in wave number space which appear to broaden, forming a spectral slope conforming to the power of $-5/3$ as they migrate (fig. \ref{2obsbest_cur_dens}).  There is one or two slightly prominent outlier modes in frequency space, but the higher wave number modes are broadband.  The spectrum in k-space conforms well to a $|k|^{-5/3}$ slope that one might expect for mode creation in a nonlinear system.  Comparing the current density fluctuations in fig. \ref{single_heat} with fig. \ref{2obsbest_obs_heat}, we conclude that the resulting flow in this arrangement of obstacles, despite a smaller Reynolds number, may be considered turbulent.

\begin{figure}[h]
	\centering
	\includegraphics[width=0.7\linewidth]{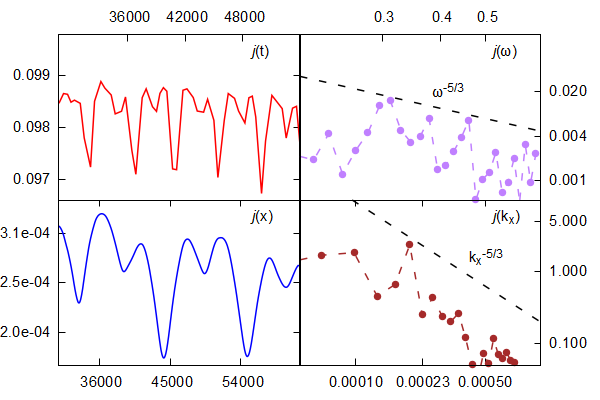}
	\caption[Current density from two close impurities]{Current density fluctuations recorded from a modeled flow obstructed by two impurities positioned relatively close.  The proximity produces a single dominant mode in wave number space (shown in log scale) similar to what is seen in $j(k_x)$ from a single obstacle (fig. \ref{single_cur_dens}), implying the effect of closely placed obstacles on the current density is similar to that caused by a single obstacle.  The multi-modal spectrum in frequency space (shown in log scale) is also similar the $j(\omega)$ spectrum for a single obstacle.  The units of the plot are indicated in Appendix \ref{appendixA}.}
	\label{2obs2close_dens}
\end{figure}

\begin{figure}[h]
	\centering 
	\includegraphics[width=0.7\linewidth]{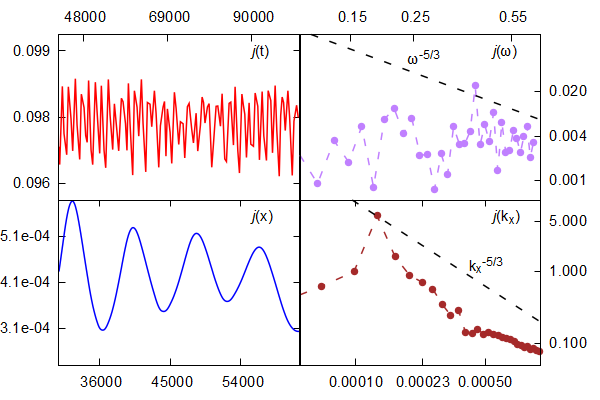}
	\caption[Current density from two distant impurities]{Current density fluctuations from a model simulating a flow around two impurities situated at a large distance relative to their diameter.  In this arrangement two discrete modes appear prominent in the wave number spectrum of $j$ shown in log scale, and multiple modes are visible in the frequency spectrum of $j$, also in log scale.  A turbulent signal is not detected for obstacles placed too far apart.  The units of the plot are indicated in Appendix \ref{appendixA}.}
	\label{2obs2far_dens}
\end{figure}

We find the presence of turbulence in a two-dimensional solid such as graphene to be sensitive to obstacle placement.  A series of subsequent tests place the second obstacle at various other distances from the first in directions both parallel and perpendicular with respect to the inflow.  The second obstacle is positioned such that the flow on the front side of the obstacle is impacted to some degree by the wake created by the first.  When the trailing obstacle is positioned within approximately $2 \mu m$ of the leading obstacle, no turbulent or preturbulent signal is detected in $j$.  At this range the current density's wave number spectrum shows a single dominant mode, and the frequency spectrum shows a few distinct prominent modes with initial signs of broadening, but is not broadband (fig. \ref{2obs2close_dens}).  The spectra are very similar to those created by a single obstacle (see fig. \ref{single_cur_dens}), implying the close proximity of the obstacles has the same effect on the current density as a single, larger obstacle.  Additionally, there is no evidence of a turbulent signal when the obstacles are situated at a large distance with respect to the characteristic length of the system.  For models examined in this work, a turbulent signal is not detected when the obstacles are separated at distances greater than $5 \mu m$.  The current density fluctuations recorded in a simulation of a model with obstacles $6.02 \mu m$ apart show a single prominent mode in wave number space but a somewhat broadband spectrum in the high frequency range in frequency space, with the exception of a single outlier mode; also similar to the current density fluctuations caused by a single obstacle (fig. \ref{2obs2far_dens}).  The contribution of the non-linear effects caused by the wake of first obstacle evidently dissipates at larger distances.


\begin{figure}[h] 
	\centering
	\includegraphics[width=0.8\linewidth]{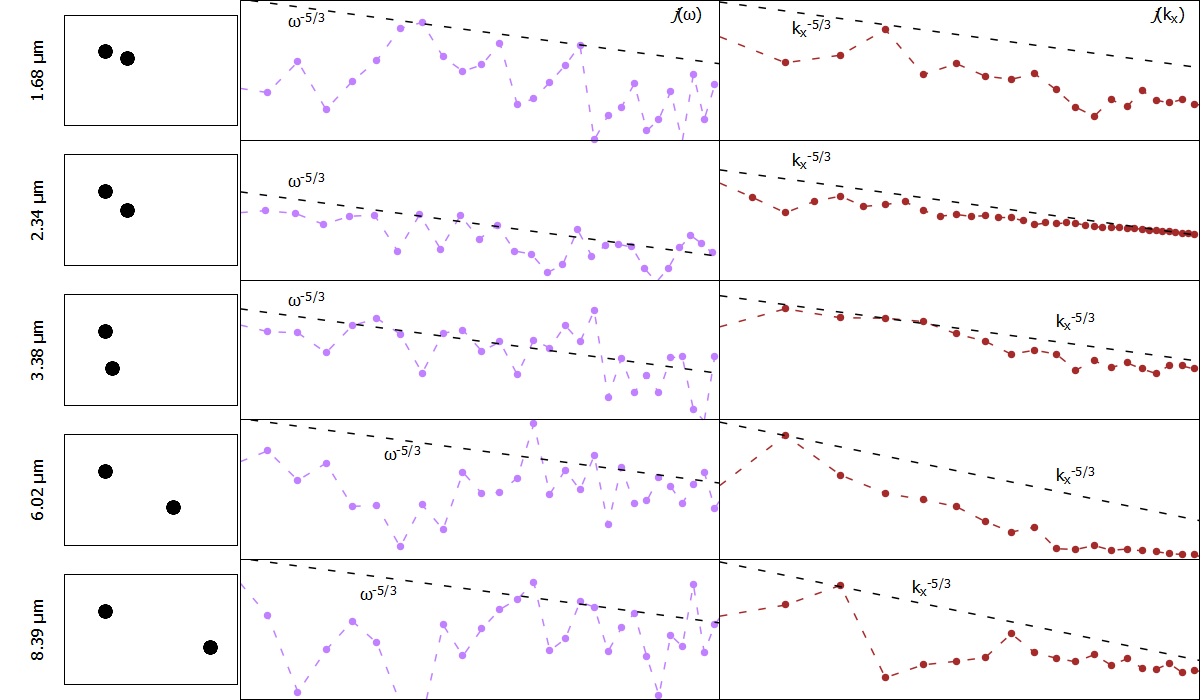}
	\caption[The effect of separation on current density]{The effect of object separation of two impurities on the current density spectra.  A turbulent signal is detected when the spectrum in both frequency space ($j(\omega)$ in log scale in the second column) and wave number space ($j(k_x)$ in the third column; log scale) are broadband.  There is no turbulent signal when the impurities are less than approximately $2 \mu m$ apart as evidenced by the lack of a broadband signal in frequency space.  A broadband signal is similarly absent when the obstacles are separated by greater than approximately $5 \mu m$. }
	\label{2obs_many_cur_dens}
\end{figure}

Fig. \ref{2obs_many_cur_dens} depicts the current density in frequency space and in k-space of five runs with increasing obstacle placement.  At the smallest distance, $1.68 \mu m$, the frequency space spectrum is multi-modal while the k-space spectrum has a single prominent mode, resembling the spectral effects of a single obstacle.  At a slightly larger distance, $2.34 \mu m$, the k-space spectrum becomes broadband and conforms to the slope of  $|k|^{-5/3}$.  More modes are present in the frequency spectrum, but it is not definitively broadband.  A broadband signal is present in both frequency space and wave number space for objects placed at a distance of $3.38 \mu m$ (also shown in fig. \ref{2obsbest_obs_heat}), indicating turbulence.  As the distance increases from $3.38 \mu m$ to $6.02 \mu m$, the frequency spectrum reverts back to multiple modes, and the broadband spectrum in k-space dissipates into a single mode.  The turbulent signal is gone when the objects are separated at this distance.  At the largest separation tested, $8.39 \mu m$, a small number of prominent modes are present in both spectra of $\bm{j}$ so that it also resembles the current density spectra created by a single obstacle.  Further tests investigating different placement configurations (not shown) indicate that detection of turbulence is sensitive to other positional features such as alignment.  Because one obstacle must be within the influence of the other's wake, when the objects are adjacent and too far apart a turbulent signal is not detected.

Note that the identification of a broadband spectrum contains an element of judgment, and a determination for the presence of turbulence using the spectrum is not rigorous.  Because each model configuration shows multiple modes in frequency space, and because the most prominent mode for each in k-space could be considered an outlier, it is possible that each depicted obstacle configuration can be considered capable of producing a turbulent signal.  Increasing the number of lattice nodes in the model will produce a higher resolution spectrum and subsequently more confidence in the interpretation, but it would also be more computationally expensive.

\section{Conclusions}

To summarize, we were able to detect a potential turbulent flow in the relativistic hydrodynamics of massless charged quasi-particles in an ultra pure, idealized sample of graphene.  Though the Reynolds number of the modeled systems were on the lower edge of the preturbulent range, evidence of turbulence emerges as a result of interactions in the wake of multiple obstacles placed in the sample.  For two similarly sized impurities a signal of turbulence is dependent upon their separation and position, while apparently independent of size or shape.  This dependence creates a more complex formulation of turbulence that cannot be represented by the Reynolds number metric alone, and has consequences for the conductivity of graphene.  The sensitivity to turbulence based on impurity placement is similar to the sensitivity of the placement of wind turbines in a wind farm.  A notable difference, though, is the importance of higher altitudes above the turbines on the air flow through the farm.  The electric flow within two-dimensional graphene is not affected by out of plane effects.  

A preturbulent signal was detected in a test modeled after \cite{PhysRevLett.106.156601}, in agreement with the findings of that work.  Further investigation is needed to study the effect of semi-rigid obstacles and momentum relaxation on the production of a turbulent signal.  The resonance of a semi-rigid obstacle caused by the flow would amplify the current density fluctuations, and the effect would likely be multiplied by the presence of additional semi-rigid obstacles.  However, the effect would be balanced or mitigated by momentum relaxation effects caused by the obstacle and the lattice itself.  The lattice structure of another two-dimensional solid such as a Kagome metal should also be investigated with the employed RLBM numerical modeler.  The two-dimensional molecular lattice of a Kagome metal creates a linear dispersion relation at low energies, similar to that of graphene, but it has a much stronger electric coupling constant that enables a faster Dirac fluid flow \cite{2019arXiv191106810D_ORIG}.  With a similar viscosity, this implies a higher likelihood of detecting a preturbulent or turbulent flow.

%
%

\chapter{Summary and Conclusions}

In Chapter \ref{stirring_force_project} we explored the properties of a massless fluid system incorporating a stirring force into a relativistic modification of the Lattice Boltzmann Method, driving the simulation of a two-dimensional fluid system at small spectral scales.  The energy spectrum of the modeled flow demonstrated the slope expected for an inverse cascade of energy propagation.  The detection of an energy cascade lends confidence that the RLBM modeler, along with the addition of a forcing term, is capable of a reasonable reproduction of turbulent flows for a fluid of massless particles.    

In Chapter \ref{obstacle_project} we found that a turbulent flow in a sample of graphene in a realistic but ideal setting is possible, although it requires the interaction between multiple impurities and is dependent upon their separation and position.  Though, it is apparently only weakly dependent on size or shape at reasonable scales.  We were able to reproduce the preturbulent flow of massless charge carriers within graphene demonstrated by \cite{PhysRevLett.106.156601}, but were able to produce a fully turbulent flow with an additional impurity incorporated within the model at specific separation distances.      

The sensitivity of turbulence to impurity placement is similar to the placement sensitivity of wind turbines in a wind farm.  An engineer must take into account the effect the wakes created by each turbine have on their neighbors when designing the arrangement of the turbines in order to minimize wind turbulence that will disrupt the farm's efficiency.  

The RLBM modeler is readily applied to other two-dimensional solid systems exhibiting a low energy linear dispersion relation such as a Kagome solid.  In a Kagome solid the massless fluid flow is believed to have a similar viscosity to graphene, but can reach larger velocities owing to a stronger electric coupling constant \cite{2019arXiv191106810D_ORIG}, and is also experimentally accessible.  Further applications of the RLBM in the exploration of turbulent flows should continue to be investigated.  An investigation of the energy cascade of a three dimensional system will help validate the model's ability to reproduce turbulence.  Different turbulence-generating forcing schemes should be considered in order to achieve higher velocities.  

The effect of semi-rigid obstacles and momentum relaxation on the production of a turbulent signal can be investigated with a modest alteration of the boundary handling in the RLBM code.  The resonance of a semi-rigid obstacle caused by the flow would amplify the current density fluctuations, and the effect would likely be multiplied by the presence of additional semi-rigid obstacles.  However, the effect would be balanced or mitigated by momentum relaxation effects caused by the obstacle and the lattice itself.  

Further examination of the turbulent flow created by two or more impurities at transitional Reynolds number ranges should be carried out to determine, in better detail, the relationship between obstacle placement and the resulting turbulence.  This might include an objective quantification of the presence of a broadband signal in the spectra of the current density and how it relates to the position of a second particle in two dimensions.  The approach should then be extended to three dimensions.  The vortex dynamics involved in a turbulent flow differs significantly between systems in two and three dimensions, thus the relationship can be expected to be very different, if it indeed exists.

Having successfully reproduced the turbulent flow of graphene's charged quasi-particles, this formulation of the Relativistic Lattice Boltzmann Method can be assumed to reasonably reproduce turbulence in relativistic flows that are less accessible to experiment.  It therefore may be used, for example, to describe the group dynamics of the particles on the outside of the event horizon of a black hole, which are thought to make up a relativistic fluid.  It can also be applied to other relativistic flows like the quark-gluon plasma that is created just after a high energy particle collision, or the relativistic jets created by compact astrophysical objects such as a neutron star or black hole.

\singlespacing

\bibliographystyle{unsrt}
\bibliography{lambda}

\appendix

%
%

\doublespacing
\chapter{Maxwellian Form of the Equilibrium Distribution Function} \label{chapter:maxwellian_feq_derivation}
\singlespacing

The form of the equilibrium probability distribution function was determined by Maxwell using logic described in \cite{doi:10.1080/14786446008642818} and is summarized as follows.  When two particles collide (not point particles) their resulting trajectories are highly sensitive to the relative angles of collision, so that when considering a large number of particles, the number of collisions results in angular distribution of the particle directions.  After a time, the particles within the fluid will reach an equilibrium so that, on average, the velocities of the particles in the fluid match the mean fluid velocity $(\bm{v} = \bm{u})$.  That is, the fluid is isotropic in velocity space in a reference frame moving with the speed of $\bm{u}$.  Then, the equilibrium distribution function is
\begin{equation} \nonumber
	f_{eq}(t, |v|, x)
\end{equation}
For this line of logic, we focus specifically on the distribution of the velocity parameter.  We assume that the distribution function is separable in velocity.
\begin{equation} \nonumber
f_{eq} (|v|^2 ) = f_{eq} (v_x^2 + v_y^2 + v_z^2 ) = f_{eq,x}( v_x^2 ) f_{eq,y} ( v_y^2 ) f_{eq,z} (v_z^2 )
\end{equation}
At equilibrium, the velocity is isotropic and constant.
\begin{equation} \nonumber
f_{eq} (|v|^2 )= C
\end{equation}
That being the case, it is also true that $ln( f_{eq} (|v|^2 ) ) = ln( C )$ and $ln(C) = C$.  Then
\begin{equation} \nonumber
ln( f_{eq,x} ( v_x^2 ) f_{eq,y} ( v_y^2 ) f_{eq,z} ( v_z^2 ) ) = C
\end{equation}
and
\begin{equation} \label{eq:ln_feq_separable}
ln( f_{eq,x} ( v_x^2 ) ) + ln( f_{eq,y} ( v_y^2 ) ) + ln( f_{eq,z} (v_z^2 ) ) = C
\end{equation}
This equation is fulfilled if each separated one-dimensional equilibrium distribution is of the form $ln(f_{eq,x}( v_x^2 )) = a + bv_x^2$, where $a$ and $b$ are both constants.  Plugging that in to \ref{eq:ln_feq_separable} we have
\begin{equation} \nonumber
	ln( f_{eq,x} ( v_x^2 ) ) + ln( f_{eq,y} ( v_y^2 ) ) + ln( f_{eq,z} (v_z^2 ) ) =
	3a + b\left( v_x^2 + v_y^2 + v_z^2 \right).
\end{equation}
This works because the right-hand side is constrained to be constant, and $a$, $b$ and $ |v| = v_x^2 + v_y^2 + v_z^2 $ are all identified as constant.
Then, the form of the equilibrium distribution function must be
\begin{equation}
	f_{eq}(|v|) = e^{3a} e^{b|v|^2},
\end{equation}
which is Maxwellian.  Assuming monatomic collisions that conserve mass, momentum and energy, we use the moment equations (\ref{eq:moments_cl}) to find the constants a and b.
\begin{equation}
	f_{eq} (x,|v|,t)
	= \rho \left( \frac{\rho}{2 \pi p} \right)^{3/2} e^{- \frac{p|v|^2}{2 \rho}}
\end{equation}

%
%

\doublespacing
\chapter{Three Dimensional Orthogonal Polynomials}  \label{appendixA}	
\singlespacing

The vector polynomials $ P_{i_1 ... i_n}^{(n)} \left( \bm{v} \right) $ are constructed through the orthogonality condition with respect to the angular integral $ \int \frac{d\Omega}{4\pi} $ \cite{2011PhRvC..84c4903R}.  The first few polynomials are 
\begin{align} \label{orthogonal_polynomials_spherical}
	\nonumber P^{(0)} =& 1, \\
	\nonumber P_i^{(1)} =& v_i, \\
	\nonumber P_{ij}^{(2)} =& v_i v_j - \frac{1}{3} \delta_{ij} , \\
	\nonumber P_{ijk}^{(3)} =& v_i v_j v_k - \frac{1}{5} \left(  \delta_{ij} v_k + \delta_{ik} v_j + \delta_{jk} v_i \right), \\
	...
\end{align}
The first few orthogonality relations are
\begin{align}
	\nonumber \int \frac{d\Omega}{4\pi} P^{(0)} P^{(0)} = & 1, \\
	\nonumber \int \frac{d\Omega}{4\pi} P_i^{(1)} P_j^{(1)} = & \frac{\delta_{ij}}{3} , \\
	\nonumber \int \frac{d\Omega}{4\pi} P_{ij}^{(2)} P_{lm}^{(2)} = & \frac{1}{15} \left( \delta_{il}\delta_{jm} + \delta_{im}\delta_{jl} - \frac{2}{3} \delta_{ij}\delta_{lm}  \right)
\end{align}

\chapter{Discrete the Forcing Term}  \label{appendixB}

The forcing term is projected onto orthogonal polynomials $P^{(n)}$ up to the second order. 
$$
F_i (\partial_p^i f) = e^{-\bar{p}} \sum_{n=0}^{2} {a^{(n)}P^{(n)} }
$$
The projection coefficients $a^{(n)}$ are dependent on position and time, and the external force $F_i$ is independent of momentum.  The divergence of the Boltzmann probability distribution function $f$ is with respect to momentum.  The time element of momentum, $\bar{p}=\frac{p^0}{T_0}$, is also the magnitude of the momentum.  We use the first three multi-dimensional orthogonal polynomials in \ref{orthogonal_polynomials_spherical}. The coefficients are obtained with the orthogonality relationship:
\begin{eqnarray}
	\nonumber
	\int_{0}^{\infty} d\bar{p} \bar{p}^2 \int{\frac{d\Omega}{4\pi} {\bar{p}} F_i \left( \partial_p^if \right) P^{(m)}} \\
	\nonumber 
	=
	\int_{0}^{\infty} d\bar{p} \bar{p}^2 & \int{\frac{d\Omega}{4\pi} {\bar{p}} \left( e^{-\bar{p}}\sum_{n}^{\infty}{a^{(n)}P^{(n)}} \right)P^{(m)}},  \\
	\nonumber
	= 
	\int_{0}^{\infty} d\bar{p} \bar{p}^2 & \int{\frac{d\Omega}{4\pi} {\bar{p}} e^{-\bar{p}}a^{(m)}\left(P^{(m)}\right)^2} .
\end{eqnarray} 

For the zero order coefficient, the orthogonality relationship is,
\begin{eqnarray}
	\nonumber
	\int_{0}^{\infty} \int{\frac{d\Omega}{4\pi} } d\bar{p}\ {\bar{p}}^3\ F_i \left(\partial_p^i f\right) P^{(0)} \\
	\nonumber
	= \int_{0}^{\infty} & \int\frac{d\Omega}{4\pi} d\bar{p}\ {\bar{p}}^3 e^{-\bar{p}} a^{(0)} \left(P^{(0)}\right)^2 .
\end{eqnarray}
The integrals are regarded as spherical volume integrals in velocity space where $\Omega$ is the solid angle, and $\bar{p}$ is the unit-less magnitude of the momentum representing the radius extending to infinity.  We obtain,
\begin{equation}
	\label{eq:expCoef01}
	a^{(0)} = - \frac{F_i}{6 T_0} \int_{V} dV v^i f .
\end{equation}
where the volume integral is represented as $\int_{V} dV = \int_{0}^{\infty} d\bar{p}\ {\bar{p}}^2 \int \frac{d\Omega}{4\pi}$.  The divergence theorem at an infinite boundary was used and the divergence of the magnitude of the momentum, given by $\partial_p^i \bar{p} = \frac{v^i}{T_0}$, was also used.  

The orthogonal expression for the first order coefficient is, 
\begin{eqnarray}
	\nonumber
	\int_{0}^{\infty} \int \frac{d\Omega}{4\pi} d\bar{p}\ {\bar{p}}^3\ F_i \left(\partial_p^if\right) P^{(1)j} \\
	\nonumber
	= \int_{0}^{\infty} \int & \frac{d\Omega}{4\pi} d\bar{p}\ {\bar{p}}^3 e^{-\bar{p}} a_k^{(1)} P^{(1)k} P^{(1)j} ,
\end{eqnarray}
\begin{eqnarray}
	\nonumber
	\int_{0}^{\infty} \int \frac{d\Omega}{4\pi} d\bar{p} {\bar{p}}^3 F_i \left(\partial_p^if\right) v^j &=& 2 a_j^{(1)} , \\
	\nonumber
	- F_j \frac{1}{T_0} \int_{V} dV f &=& 2 a_j^{(1)} , \\
	\label{eq:forceExpCoef00}
	a_j^{(1)} &=&  - \frac{1}{2} \frac{F^j}{T_0} \int_{V} dV f .
\end{eqnarray}
Here we evaluated the angular portion of the integral with the relationship $\int \frac{d\Omega}{4\pi} v^i v^j = \frac{1}{3} \delta_{ij}$, and we used the divergence of the velocity relation $\partial_p^i v^j = \frac{1}{p^0} ( \delta^{ij} - v^i v^j ) $.

Finally, we used the angular integral $\int \frac{d\Omega}{4\pi} v^i v^j v^l v^m = \frac{1}{15} \left( \delta_{il}\delta_{jm} + \delta_{im}\delta_{jl} + \delta_{ij}\delta_{lm} \right) $ to determine the second order projection coefficient.
\begin{eqnarray}
	\nonumber
	\int_{0}^{\infty} \int \frac{d\Omega}{4\pi} d\bar{p}\ {\bar{p}}^3\ F_k \left( \partial_p^k f\right ) P^{(2)lm} \\
	\nonumber
	=
	\int_{0}^{\infty} \int \frac{d\Omega}{4\pi} d\bar{p}\ {\bar{p}}^3 e^{-\bar{p}} a_{ij}^{(2)} & P^{(2)ij} P^{(2)lm} , 
\end{eqnarray}
\begin{eqnarray}
	\nonumber
	a_{ij}^{(2)} \frac{4}{5} & = &
	F_k \frac{1}{T_0} \int_{V}dV f v^k v^i v^j 
	- 2 F_i \frac{1}{T_0} \int_{V}dV f v^j \\ \nonumber
	& \ & +\frac{1}{3} \delta^{ij} F_k \frac{1}{T_0} \int_{V} dV f v^k    
\end{eqnarray}

For this calculation we note that the trace of the velocity is unity, containing only angular information.  Also, the second order orthogonal polynomial is trace-less, as is its coefficient  $a_{ij}^{(2)}$, so that $ a_{ij}^{(2)} \delta_{ij} = 0 $.  Then the second order projection coefficient is,

\begin{equation}
	\label{eq:forceExpCoef02}
	a_{ij}^{(2)}  
	=  \frac{5}{4}  F_k \frac{1}{T_0} \int_{V}dV f v^k v^i v^j 
	-  \frac{5}{2}  F_i \frac{1}{T_0} \int_{V}dV f v^j 
	+ \frac{5}{12} F_k \frac{1}{T_0} \delta^{ij} \int_{V} dV f v^k .
\end{equation}

The forcing term projected onto orthogonal polynomials up to the second order is then, 
\begin{align}
	\nonumber
	F_i \partial_p^i f 
	= e^{-\bar{p}}  \frac{1}{T_0}  & \left[
	\frac{3}{12} F_i \int_{V}dV v^i f  
	-         \frac{1}{2}  F^i v^i \int_{V} dV f \right. \\ 
	&	+ \left.  \frac{5}{4}  F_i v_j v_k \int_{V}dV f v^i v^j v^k
	-         \frac{5}{2}  F_i v_i v_k \int_{V}dV f v^k 
	\right] + \ensuremath{\mathcal{O}}\left( v^3 \right).
\end{align}

\chapter{Minimum Simulation Time}  \label{sec:minSimTime}

\begin{figure}
	\centering
	\includegraphics[width=1.0\linewidth]{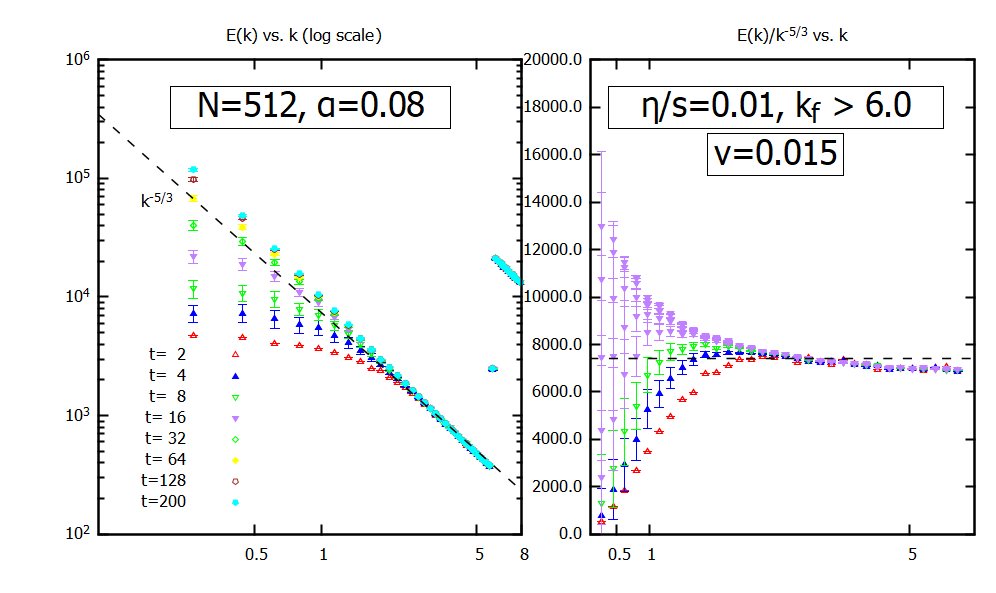}
	\caption[Development of Energy spectrum in time]{Power spectrum for simulations with energy injected at $k_f = 6.0$ at driving force magnitude $|F|=800$ for simulation durations $t=2$ through $200$.  A general conformity to the Kolmogorov spectrum is obtained after $t=32$.}
	\label{fig:mintimedualplot}
\end{figure}

Convergence to a stable energy spectrum for a simulation was achieved before $32$ simulated time units.  Figure (\ref{fig:mintimedualplot}) shows the power spectrum for a simulation at runtime $t=2$ through $t=200$ time units at step sizes of roughly $2^2$.  It is the same power spectrum plot as in \ref{stirring_force_project}, but represents a single simulation instead of the power spectrum averaged over a number of executions. It indicates a stable spectral slope is realized after $32$ time units and remains stable to large time windows.  The companion plot depicts the same values divided by the predicted slope providing a higher resolution picture.  The plot depicts large errors at the large scales; a result of numerical limitations for small wave numbers in a discrete spectrum.  However, the error for the spectrum with the shorter time span engulfs the spectrum for longer time spans, displaying an evolving stability.

\begin{figure}[h] 
	\centering
	\includegraphics[width=1.0\linewidth]{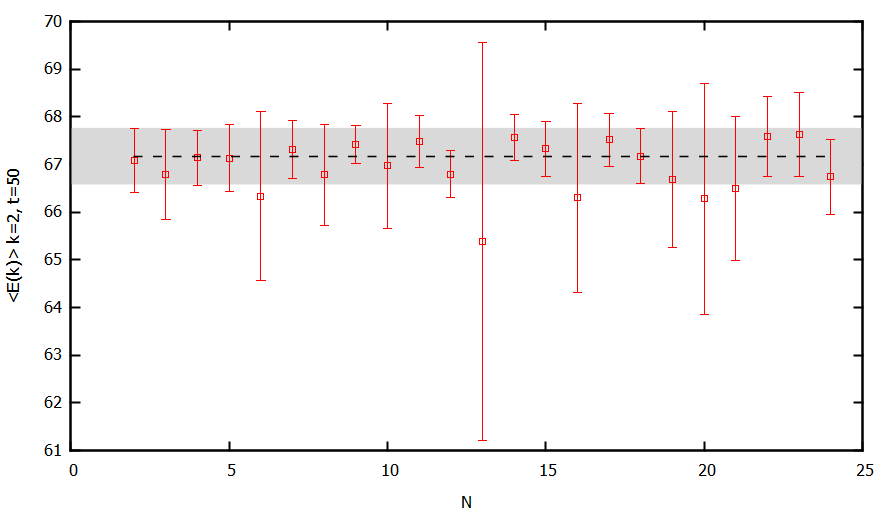}
	\caption[Determination of the time step]{The average Power Spectrum $ \langle E(k) \rangle $ vs. $N$, where $N=\frac{t}{\Delta t}$ ($\Delta t = 0.05$) is a range of candidate time steps to best mitigate autocorrelation risk.  The gray band indicates agreement within the margin of error of the $N=18$ time step.  The power spectrum is for $k = 2.0$ in a simulation with energy injected at scales greater than $4.0$ at maximum driving force magnitude of $400$.  The simulation was run for $t=50$ time units. }
	\label{fig:nvsEkwithErrCT}
\end{figure}

The error is calculated as $ \frac{ 1 }{N} \sum_{i=1}^{N} \sqrt{ \langle E(k) \rangle^2-E_N(k)^2}$ where $ \langle E(k) \rangle $ is a running average in time for each $k$.  Here $N$ is a time step determined by a suitable disassociation time $t$ where $N=\frac{t}{\Delta t}$ ($ \Delta t = 0.05 $).  At a smaller time step the average poses an autocorrelation danger.  So, a suitably large $N$ was determined to remove the autocorrelation risk and maximize the number of time steps participating in the average.  Figure (\ref{fig:nvsEkwithErrCT}) shows $\langle E\left(k\right) \rangle $ for $k_f=2.0$ at $t=50$, plotted against candidate values of $N$, with the error in red.  $N = 18$ (and therefore a disassociation time of $t=0.9$) was determined to be acceptably large because of its agreement, within the margin of error, with all other values of $N$ (as shown by the gray band), and because of its relatively small error.  Many other choices of $N$, including $4$ or $5$, could have also been chosen to achieve the same result.  This analysis is not statistically rigid, but suitable enough for the purpose.

\pagebreak

\doublespacing
\chapter{Determination of the Equation of State} \label{finding_eq_st}
\singlespacing

The employed equation of state, $p = \frac{\epsilon}{3}$, that of an ideal (three-dimensional) gas, is easily determined by considering the stress-energy tensor of a fluid in a state of equilibrium, where the fluid is at rest in its reference frame; $u^i = 0$.
\begin{equation} \label{equation:stress-energy-tensor-equilibrium}
	T^{\mu\nu} = \left( \begin{matrix}
		\epsilon &  0 &  0 &  0 \\
		0        & -p &  0 &  0 \\
		0        &  0 & -p &  0 \\
		0        &  0 &  0 & -p
	\end{matrix} \right).
\end{equation}
The stress-energy tensor is both scale and conformally invariant, so its trace is $0$ \cite{romatschke2010new}.  The zero trace is evident when the energy density and pressure of an ideal relativistic gas in equilibrium are given in terms of the equilibrium distribution function and the momentum and fluid velocity 4-vectors ($P^\mu$ and $U^\mu$ respectively)  \cite{romatschke2012relativistic}. 
\begin{align}
	\nonumber \epsilon = \int d\chi \left( P^\mu U_\mu \right)^2 f_{eq} , \\
	\nonumber p = - \frac{1}{3} \int d\chi \left[ \left(  P^2 - P^\mu U_\mu \right)^2 \right]  f_{eq}
\end{align}
The differential in the integrals is defined in \cite{romatschke2012relativistic}, but its definition is not necessary for the calculation.  The trace of the stress-energy tensor of an ideal relativistic gas in equilibrium is then written as
\begin{align} 
	\nonumber Tr( T^{\mu\nu} ) &= \int d\chi \left[
		\left( P^\mu U_\mu \right)^2 + \frac{3}{3}  P^2 + \frac{3}{3} \left( P^\mu U_\mu \right)^2  
	\right] f_{eq}, \\
	\nonumber  &= \int d\chi \left[
	   P^2 
	\right] f_{eq}.
\end{align}
For a massless particle $E=|\bm{p}|$, where $\bm{p}$ is the momentum 3-vector.  Therefore, $P^2 = E^2 - \bm{p}^2 = \bm{p}^2 - \bm{p}^2 = 0$, and consequently $Tr( T^{\mu\nu} )=0$.  Then the equation of state is $p = \epsilon/3$.  

Note that the relativistic lattice Boltzmann model employed in this work uses a two dimensional spatial lattice, but a three-dimensional momentum lattice.  Its moments are derived from the $3 + 1$ stress-energy tensor, so the three-dimensional equation of state is appropriate. 

\chapter{Computational Platform}

All simulations were executed on the Eridanus high availability computing cluster at the University of Colorado in Boulder, Colorado.  The cluster provides performant distributed execution of simulations through PBS (Portable Batch System) services using the Torque resource manager.  It consists of 28 computing nodes, each with 12 cores and 32 GB of memory.

\end{document}